\documentclass[mathematics,article,accept,moreauthors,pdftex]{Definitions/mdpi}

\firstpage{1} 
\pubvolume{1}
\issuenum{1}
\articlenumber{0}
\pubyear{2021}
\copyrightyear{2021}
 \externaleditor{Academic Editor(s):{Josue Antonio Nescolarde Selva}}
\datereceived{2 January 2021} 
\dateaccepted{18 February 2021} 
\datepublished{} 
\hreflink{https://doi.org/10.3390/}

\usepackage{graphicx}
\usepackage[utf8]{inputenc}
\usepackage{graphicx}
\usepackage{amsfonts}
\usepackage{amsmath}
\usepackage{multirow}
\usepackage{pstricks}
\usepackage{pgf,tikz}
\usepackage{algorithmic}
\usepackage{algorithm}
\usepackage{booktabs} 
\usepackage{multirow}
\usepackage{soul} 
\usepackage{microtype}

\setitemize{parsep=6pt,itemsep=0pt,leftmargin=*,labelsep=5.5mm,align=parleft}
\setenumerate{parsep=6pt,itemsep=0pt,leftmargin=*,labelsep=5.5mm,align=parleft}
\usepackage{comment}
\usepackage{verbatim}
\usepackage{mathrsfs}
\usetikzlibrary{arrows}
\usepackage{caption}
\usepackage{rawfonts}
\usepackage{pictexwd}
\usepackage{subcaption}
\graphicspath{{Images/}}
 \usepackage[labelformat=simple]{subcaption}

\DeclareCaptionLabelFormat{subcaptionlabel}{\normalfont(\textbf{#2}\normalfont)}
\Title{Community Detection Problem Based on Polarization Measures: An Application to Twitter: The COVID-19 Case in Spain}
\TitleCitation{Community Detection Problem Based on Polarization Measures: An Application to Twitter: The COVID-19 Case in Spain}

\Author{{Inmaculada}~{Gutiérrez}~ $^{1,}$*\orcidA{}, Juan Antonio {Guevara} $^{1}$\orcidB{}, {Daniel}~Gómez~ $^{1,2}$\orcidC{}, Javier {Castro} $^{1,2}$\orcidD{} and~{Rosa}~Espínola~$^{1,2}$\orcidE{} } 

\AuthorNames{Inmaculada  Gutiérrez, Juan Antonio Guevara, Daniel Gómez,  Javier  Castro and Rosa Espínola}
\AuthorCitation{Gutiérrez, I.; Guevara, J.A.;  Gómez, D.; Castro, J.; and Espínola, R.}
\address{%
$^{1}$ \quad Faculty of Statistics, Complutense University 
Puerta de Hierro, s/n, 
  28040 Madrid, Spain; inmaguti@ucm.es ~(I.G.); juanguev@ucm.es~(J.A.G.); dagomez@ucm.es (D.G.); jcastroc@ucm.es (J.C.); rosaev@ucm.es (R.E.)\\
$^{2}$ \quad Instituto de Evaluaci\'{o}n Sanitaria, Complutense University, {28040} Madrid, Spain }

\corres{Correspondence: inmaguti@ucm.es}

\abstract{In this paper, we address one of the most important topics in the field of Social Networks Analysis: the community detection problem with additional information. That additional information is modeled by a fuzzy measure that represents the risk of polarization. 
Particularly, we are interested in dealing with the problem of taking into account the polarization of nodes in the community detection problem.  Adding this type of information to the community detection problem makes it more realistic, as a community is more likely to be defined if the corresponding elements are willing to maintain a peaceful dialogue. The polarization capacity is modeled by a   fuzzy measure  based on the $JDJ_{pol}$ measure of polarization related to two poles. We also present an efficient algorithm for finding groups whose elements are no polarized. Hereafter, we work in a real case. It is a network  obtained from Twitter, concerning the political position against the Spanish government taken by  several influential users. We analyze how the partitions obtained change when some additional information related to how polarized that society is, is added to the problem.}


\begin{document}

\section{Introduction}\label{sec:intro}
    The field of Social Networks Analysis (SNA) encompasses a wide range of processes devoted to the investigation of social structures modeled by complex networks or graphs. These are models to show  schemes of relations between the entities of a complex system, be it in technological applications, nature, or society, so that the elements of the systems  are described as vertices or nodes, and their interactions as  links or edges. Particularly, online social networks are usually represented by a graph whose nodes are people and whose edges show relations of different nature: social, friendship, common interests, familiarities, etc. Thousands of millions of data are constantly generated, so the importance of the SNA has grown more and more in the recent decades, attracting the interests of many researchers from different areas.  Generally, three analysis levels can be distinguished in SNA processes: the first, related to  individuals; the next, related to the structures and relationships established by the graph structure; and the last, related to the analysis of interactions between previous levels.

One of the features shown by complex networks is their internal group structure, a property which is far from being trivial. Trying to find these structures has become a highly relevant study topic in the SNA field: the well-known community detection problem.
This problem has evolved into an essential one, having many different applications in several areas {such as biology, sociology, Big Data, or pattern recognition \cite{bennett,chaker,harakawa,tamura}.  }
From the knowledge of the community structure of a complex network, several non-trivial internal features or organizations can be reached. Furthermore, it facilitates a better understanding of the dynamic processes which take place in the network and the inference of some properties or interactions between the elements.

Therefore, the main goal of applying community detection algorithms to online social networks is to group individuals---represented by nodes---into communities, with the intention of knowing the internal structure of a given society. In this light, community detection and social polarization are closely related. In broader terms, polarization can be understood as the split of a given population into two opposite groups, both with significant and similar size. Polarization measures \cite{ER94,guevara2020measuring} provide a single value which shows all these characteristics, taking into consideration some knowledge about the similarity between the individuals, the clusters in their population, etc. Thus, the structure of a given set of individuals impacts the polarization values shown by a given measure, as well as the presence of polarization---or its degree---determines the topological structure of a network. 

 Because of the growing importance of the community detection problem, an extensive range of methods have been proposed to solve it
\cite{clauset, large, biological}, among which it is worth  highlighting the Louvain \cite{blondel} algorithm. {It is a fast multi-phase method  based on local moving and modularity optimization \cite{smart,girvanNewman}}, which provides good quality non-hierarchical partitions of the set of nodes, without a priori knowledge of the number of communities. 
 {Almost all the methods found in the literature have a point in common: the search of groups is based on the structural or topological characteristics of the graph \cite{fortunato}. Particularly, the Louvain algorithm focuses on the edges between the nodes.}
In this vein, the only information considered for the definition of groups is the knowledge represented by the graph, without deeming any additional data.
Going a step further, several authors agree on the idea of adding additional information to the graph, {be it in a Game Theory context \cite{cohesiveness,centralityPower}, by considering fuzzy sets \cite{Devarajan} or fuzzy graphs \cite{Nair2007}.} On our part, we consider the inclusion of some knowledge about the polarization of the elements of a graph when grouping them. We agree on the importance of having groups whose elements are willing to peaceful dialogue, so that they  are not prone to conflict.

\begin{figure}[H]
\begin{subfigure}{0.48\linewidth}
    \includegraphics[scale=0.26]{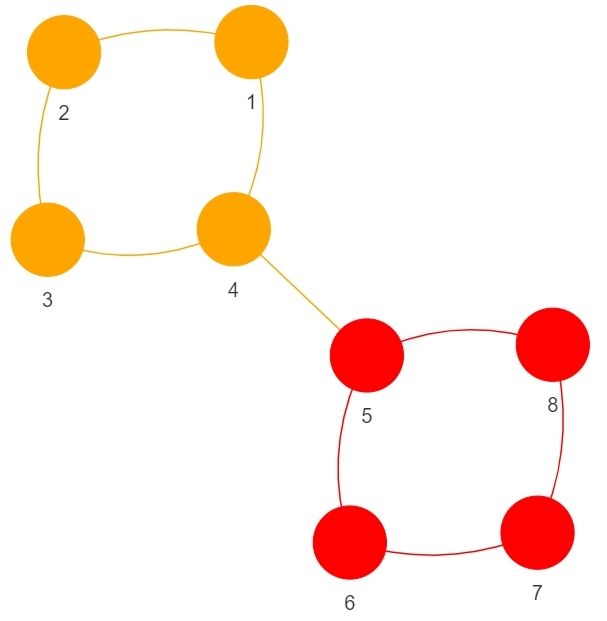}
    \caption{Classical algorithm partition.}
        \label{fig:Toy_1}
\end{subfigure}
    \begin{subfigure}{0.48\linewidth}

    \includegraphics[scale=0.26]{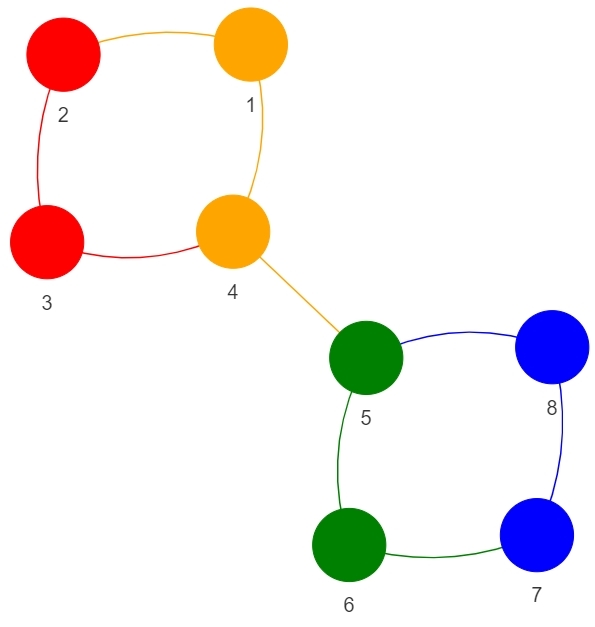}
    \caption{Extra-information algorithm partition.}
    \label{fig:toy_2}
\end{subfigure}\caption{Toy example: a graph with $8$ nodes.}\label{fig:toy}
\end{figure}

Let us illustrate this idea with a basic example. {Let the graph or network be $G=\left(V,E\right)$, consisting of a set $V$  with $|V|=8$ nodes by the edges of the set $E$   as it is shown in \mbox{Figure \ref{fig:toy}}.}  Every community detection algorithm based on the graph structure, particularly on modularity optimization \cite{girvanNewman,blondel}, organizes the elements into two clusters with $4$ each one, by separating the two wheels, i.e., $P=\{\{1,2,3,4\},\{5,6,7,8\}\}$. 
However, let us assume some knowledge about political position of the individuals against a government, represented by the vector $O=\left(+,-,-,+,+,+,-,-\right)$, so that if $O_i=+$, the individual $i$ is in favor of the government, and the opposite happens when $O_i=-$. It is fair to accept that people who hold a similar political ideology, are less prone to conflict with each other than those who have opposite ideas. On this assumption, a desirable partition could be $P^a=\{\{1,4\}, \{2,3\}, \{5,6\}, \{7,8\}\}$.

Concerning this idea of adding some additional information to the community detection problem, and based on methodology proposed by Gutierrez et al. \cite{infus,ampliinfus,multipleBip} about the the community detection in graphs based on fuzzy measures, in this study we define a new method with the purpose of adding extra-information related to polarization to the community detection problem in graphs. That additional information is based on the values given by the polarization measure developed by Guevara et al. \cite{guevara2020measuring}. Our goal is to build a model consisting of a network in combination with a polarization fuzzy measure whose structure fixes properly the reality. It is the polarization extended fuzzy graph, which takes both the attitude of the people and structure characteristics of the social network into consideration. On the basis of this model, we define a community detection method, which, by having a graph and knowing the membership degree of every individual to two poles, provides realistic partitions of reality. The choice about several aggregation operators plays an essential role in this method, as it will be shown in the following pages.

The remainder of the paper is organized as follows. In Section \ref{sec:prel}, we set the basis of the paper, by introducing several concepts related to Graph Theory, fuzzy measures study, and Polarization tools. Then, in  Section \ref{sec:networksPol} we work in the definition of a new fuzzy measure based on a Polarization measure. In combination with a graph, this fuzzy measure sets the definition of a new tool: the polarization extended fuzzy graph. In parallel, we define the non-polarization fuzzy measure, to represent the capacity of a set of elements to peacefully dialogue. From this non-polarization fuzzy measure and a crisp graph, we define the non-polarization extended fuzzy graph, for which we suggest a particular application in Section \ref{sec:CDPPolari}, related to searching partitions on it. We show the performance of this new methodology in a real case, working in the detection of groups in a polarization extended fuzzy graph whose origin is Twitter. The experiment design  and the methodology can be found in Section \ref{sec:realcase}. We finish this paper in Sections \ref{sec:discussion} and \ref{sec:conclusions}, showing some discussion and conclusions about the work done.



\section{Preliminaries}\label{sec:prel}
In this section, we introduce several concepts on which this paper is based. We divide it into two main parts: one is related to networks and graphs as well as the community detection problem, and the other is related to Polarization background.

\subsection{Graphs, Fuzzy Graphs, and Extended Fuzzy Graphs}\label{subsec:graphs}
{Let us consider the crisp graph $G=\left(V,E\right)$, whose adjacency matrix is $A$, which represents the direct connections between the nodes. Beyond the classical concept of crisp graph, Rosenfeld introduced the fuzzy graphs \cite{rosenfeldFG} based on the fuzzy relations among the individuals \cite{zadeh1965}. This tool, very useful to model situations in which there is some vagueness or uncertainty about the representation of the knowledge,  has been widely used in many fields \cite{yaqoob,picturefuzzy}. Nevertheless, from a mathematical point of view, there are some situations in which fuzzy graphs may be understood as a kind of weighted graphs \cite{mordeson}. An amplified vision of this model was introduced in \cite{infus} by defining the extended fuzzy graph, a concept based on fuzzy measures.}
As it is pointed in \cite{beliakov20}, fuzzy/capacity measures are fundamental in modeling dependencies among the inputs,  and constitute a natural tool for modeling in multiple criteria decision analysis, aggregation, group decision-making, or game theory.

\begin{Definition}[Fuzzy Measure \cite{sugeno}]\label{def:fuzzMeasure} Let $V$ denote a non empty set. A fuzzy measure is a set function  $\mu: \ 2^V \longrightarrow [0,1]$ for which the following holds.
\begin{itemize}
\item $\mu(\emptyset)=0 $
\item $\mu(V)=1 $
\item $\mu(A) \leq \mu(B)$,  $ $ $\forall A, B\subseteq V$ such that $ A \subseteq B$
\end{itemize}
\end{Definition}

  Then, with the combination of the ability of the graph to model connections between elements, and the ability of the fuzzy measures to handle the capacity related to any set of elements, it was defined the extended fuzzy graph. This tool is a graph together with a fuzzy measure defined over the set of nodes. The incorporation of a fuzzy measure goes far from the previous notion of fuzzy graphs, which are limited to the consideration of pairs of elements. In this vein, by means of a fuzzy measure defined over the set of nodes, we can represent situations in which more than two nodes are implied, independent of the way they are connected through the graph. It is obvious that the representation ability of the extended fuzzy graph goes fFar from that of the existing tools, so that much more complex situations can be addressed, with a proper modeling of reality.

\begin{Definition}[Extended fuzzy graph \cite{infus}]\label{def:extFuzzyGraph} Let $G=(V,E)$ denote a graph, and let $ \mu: 2^V     \longrightarrow  [0,1]$ denote a fuzzy measure defined over the set of nodes $V$.  An extended fuzzy graph is a triplet $ \widetilde{G} = (V,E,\mu)$, also called crisp graph with fuzzy measure $\mu$.
\end{Definition}

\begin{figure}[H]
\widefigure
\begin{subfigure}{0.39\linewidth}

   \includegraphics[scale=0.2]{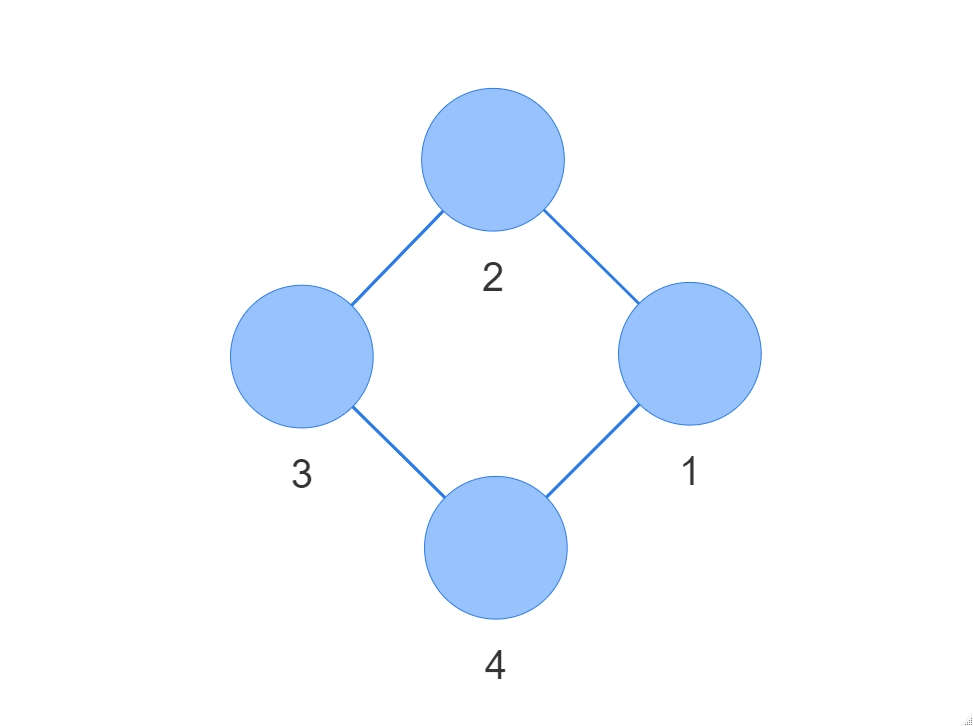}
\end{subfigure}
\begin{subfigure}{0.52\linewidth}
$$
\mu(S)=
\begin{cases}
0 & \textup{ if }  S=\emptyset\\
0 & \textup{ if }  S=\{1\},\{2\},\{3\},\{4\} \\
0  & \textup{ if }  S=\{1,2\},\{1,4\},\{2,3\},\{3,4\}\\
0.5  & \textup{ if }  S=\{1,3\},\{2,4\}\\
0.5 & \textup{ if }   S=\{1,2,3\},\{1,2,4\},\{1,3,4\},\{2,3,4\}\\
1 & \textup{ if }  S=\{1,2,3,4\}\\
\end{cases}
$$\end{subfigure}  \caption{Extended fuzzy graph $\widetilde{G}=\left(V,E,\mu\right)$.}   
    \label{fig:toyEFG}
    \end{figure}

In the following example, we show how it is possible to represent complex situations with several information sources by means of an extended fuzzy graph.
\begin{Example}
Let us consider the graph $G=\left(V,E\right)$ with $4$ nodes (Figure \ref{fig:toyEFG}). We assume some knowledge about the political position of the individuals against a government, represented by the vector $O=\left(+,-,+,-\right)$, so that if $O_i=`+'$, the individual $i$ is in favour of the government, and the opposite happens if $O_i=`-'$. These are strong political opinions, so it is not easy for individuals with opposite ideas to peaceful dialogue. However, when two individuals with the same idea are together, they can discuss peacefully at great length. Let the fuzzy measure $\mu:2^V\rightarrow [0,1]$ represent somehow the capacity of each feasible group of elements to discuss depending on their ideology. With the extended fuzzy graph     $\widetilde{G}=\left(V,E,\mu\right)$ we represent the connections between the individuals as well as their ability to peaceful dialogue regarding their political ideas. 
\end{Example}


\nointerlineskip
\switchcolumn
\vspace{-6pt}

\subsection{Community Detection Problem}\label{subsec:CDP}
One of the main applications of graphs is related to the community detection problem. Many complex networks usually have an intern modular structure so that the nodes are organized into modules with dense internal connections, scarcely interconnected externally. The goal of community detection problem is to find these hidden structures, i.e., to establish a \textit{good} partition of the set of nodes.

Some authors understand the community detection problem as an optimization problem \cite{newman2006}. 
The modularity $Q$ is one of the most used measures as objective function to be optimized. This measure, whose value is determined by the topology of the network, is used to quantify the goodness of a partition. It was first defined by Newman and Girvan~\cite{girvanNewman} and {it is usually denoted by $Q$.}

Many approaches have been proposed in the last decades to face the community detection problem \cite{newman2006,clauset, large, biological}. It is worth highlighting the Louvain algorithm \cite{blondel}, one of the most popular methods in this field, proposed by Blondel et al. in 2008. This algorithm performs very well, particularly with large networks, for which good quality non-hierarchical partitions are detected in a very little computing time. It is an iterative multi-phase method, based on modularity optimization and local moving \cite{smart}, for which the variation of modularity, $\Delta Q_i(j)$ defined in \cite{blondel}, is a key element. This variation represents the gain attained if the node $i$ is moved to the community to which $j$ belongs, and it is calculated in each step of the Louvain algorithm until a maximum of modularity is reached. The Louvain algorithm is a key point of the methodology which will be proposed in the following pages, related to the community detection in graphs with additional information.

\begin{Example}\label{ex:toyModularity}
Let us recall the graph of $8$ nodes presented in the introduction (Figure \ref{fig:toy}). There we affirmed that if the aim is to maximize the modularity, the partition should be $P=\{\{1,2,3,4\},$ $\{5,6,7,8\}\}$ (particularly, this partition is obtained with the Louvain algorithm). Observe that, indeed, $Q(P)=0.3889>0.1914=Q(P^a)$, where $P^a=\{\{1,4\}, \{2,3\}, \{5,6\}, \{7,8\}\}$ is a desirable partition that could be obtained if the additional information defined by that vector $O$ were considered.
\end{Example}

\subsection{Polarization}\label{subsec:pol}
In the last few decades, both the concept of Polarization and its measurement have aroused increasing interest in the literature. Due to the new digital technologies, Web 2.0, and social big data analysis, the study of the social conflict is now more reachable than ever. In broader terms, Polarization can be understood as the split of a given society into two different and opposite groups along an attitudinal axis. The measurement of the Polarization is studied in several disciplines \cite{ER94, reynal2001ethnic,Apouey_2007,permanyer2015measuring}. 
{In this work, we recall the concept of Polarization based on fuzzy sets developed in \cite{guevara2020measuring}.} Guevara et al. introduced a Polarization measure based on the fuzzy set approach, with which it is possible to avoid the duality Yes/No. Due to the fuzzy sets nature, this measure can deal with numeric, ordinal, or linguistic variables as well. The main argumentation of that work is based on the assumption that ``reality is not black and white''. When considering the classical Polarization measures found in the literature, each individual is forced to belong to a specific position along the Polarization axis \cite{ER94}. 

In \cite{guevara2020measuring}, Aggregation Operators (AO) \cite{zadeh1965} are used to aggregate the information. AO were originally defined to aggregate the resulting values of the membership functions of a fuzzy set. Particularly, overlap functions \cite{gomez2016n} are used in this measure to show the degree $z$ of the intersection of both classes with respect to the object $c$. On the opposite, grouping functions \cite{bustince2011grouping} are used  to get the degree $z$ up to which the combination of these classes is supported. Let us detail the characterization of the $JDJ_{pol}$ measure.

We consider the finite set $V$ and the one-dimensional variable $X$  (ordinal or numeric). We assume that $X$ has two extreme and opposite values or poles: $X_A$ and $X_B$. Then, regarding the value of each element of $V$ on $X$, we can measure their membership degree to each of these poles $X_A$ and $X_B$. 

These membership degrees are represented by the membership functions $\eta_{X_A},\eta_{X_B}: V \longrightarrow [0,1]$, so that for every $i \in V$, $\eta_{X_A}(i)$ and $\eta_{X_B}(i)$ represent the membership degree of the element $i$ to the extreme pole $X_A$ and to the extreme pole $X_B$, respectively.

\begin{figure}[H]
\includegraphics[scale=0.4]{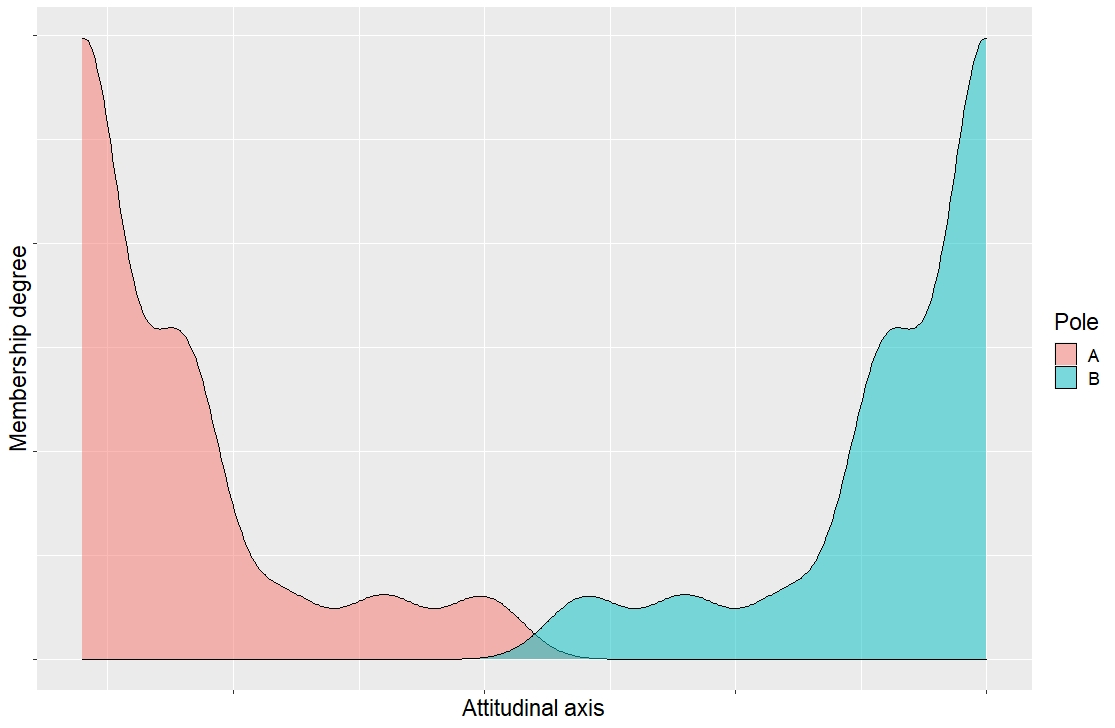}
\caption{Example of bi-polarization.}
\label{fig:bi_pol}
\end{figure}

In this scenario, Polarization exists when almost half the population is placed by the extreme position $X_A$, and the other half is placed by the extreme position $X_B$. In \mbox{Figure  \ref{fig:bi_pol}}, we show an illustrative example of two membership degree functions $\eta_{X_A}$ and $\eta_{X_B}$.

In this context, in \cite{guevara2020measuring} the $JDJ_{pol}$ polarization measure was defined as the expected risk of polarization of a given population. In this vein, $JDJ_{pol}$ measures the risk of polarization for each pair of individuals $\{i,j\}$  of a finite set $V$. The obtained value  is given by the summation of all those comparisons between all pairs and its aggregation. This value depends on the following.

\begin{itemize}
\item The closeness of the element $i$ to the pole $X_A$ and the closeness of the node $j$ to the pole $X_B$, represented by $\eta_{X_A}(i)$ and $\eta_{X_B}(j)$, respectively.
\item The closeness of the element $j$ to the pole $X_A$ and the closeness of the node $i$ to the pole $X_B$, represented by $\eta_{X_A}(j)$ and $\eta_{X_B}(i)$, respectively.
\item The grouping operator $\varphi$ chosen.
\item The overlapping operator $\phi$ chosen.
\end{itemize}

Let us provide a mathematical definition of the $JDJ_{pol}$ measure.

\begin{Definition}[$JDJ_{pol}$ Polarization measure \cite{guevara2020measuring}]\label{def:JDJ}

 Let $V$ denote a finite set, and let $\eta_{X_A}$ and $\eta_{X_B}$ denote the membership functions of the elements of $V$ to the extreme poles $X_A$ and $X_B$.
 Let $\varphi: [0,1]^{2} \rightarrow [0,1]$ denote a grouping operator and $\phi:[0,1]^{2} \rightarrow [0,1] $ denote an overlapping operator.
  Then, $JDJ_{pol}$ measure is defined as
 \begin{equation}\label{eq:JDJ}
    JDJ_{pol} (V,\eta_{X_A}, \eta_{X_B}, \varphi, \phi)= \sum_{\substack{i,j \in V \\ i \leq j}} \varphi \left( \phi ( \eta_{X_A}(i),\eta_{X_B}(j) ) ,\phi (\eta_{X_B}(i),\eta_{X_A}(j) ) \right)
\end{equation}

\end{Definition}

\begin{Remark}
The membership degrees defined by the membership function are always non-negative, particularly those degrees concerning $\eta_{X_A}$ and $\eta_{X_B}$. Then, because of the properties of the grouping and overlapping operators, the measure $JDJ_{pol}$ is monotone and non-negative. 
\end{Remark}

The performance of $JDJ_{pol}$ shows the highest values of Polarization in those cases in which the $50\%$ of the elements are located in one extreme value of the attitudinal axis and the other $50\%$ of the elements are located in the opposite extreme value. This situation is explained in detail in \cite{guevara2020measuring}.
\section{Networks with Additional Information: The Polarization Extended Fuzzy Graph}\label{sec:networksPol}
{On the basis of the Polarization measure introduced in defined in \cite{guevara2020measuring}, in this section we define a new fuzzy measure.} 
We work in the idea of adding some additional information to a given graph. To carry on with it, we assume the existence of a crisp graph $G=(V,E)$ and some knowledge  about the attitude of the elements of $V$ concerning any particular issue. First, we define a fuzzy measure emanated from the Polarization measure relative to this attitudinal knowledge. This fuzzy measure reveals the fuzzy relations existing between all the pairs of elements of $V$ emanated from the measure $JDJ_{pol}$ \cite{guevara2020measuring}. 

\begin{Definition}[Polarization fuzzy measure $\mu_{P^-}$]\label{def:muP}
Given a unidimensional variable, let $V$ denote a set of $n$ individuals, about which we know the membership degree to the extreme poles of that variable, $X_A$ and $X_B$, represented by the membership functions $\eta_{X_A}$ and $\eta_{X_B}$, respectively.
 Let the functions $\phi:[0,1]^2\rightarrow [0,1]$ $ $ and $ $ $\varphi:[0,1]^2\rightarrow [0,1]$ denote a grouping operator \cite{bustince2011grouping} and an overlapping operator \cite{gomez2016n}, respectively. 
Let $S$ denote a subset of $V$, and let $JDJ_{pol}\left(\{i,j\},\eta_{X_A}, \eta_{X_B}, \varphi, \phi\right)=\varphi(\phi(\eta_{X_A}(i), \eta_{X_B}(j)), \phi(\eta_{X_A}(j), \eta_{X_B}(i))$, according to the Equation (\ref{eq:JDJ}). We define the polarization fuzzy measure $\mu_{P^-}$ as
 
 \begin{equation}\label{eq:mup}
     \mu_{P^-}\left(S\right) = \frac{JDJ_{pol}\left(S,\eta_{X_A}, \eta_{X_B}, \varphi, \phi\right)}{ JDJ_{pol}\left(V,\eta_{X_A}, \eta_{X_B}, \varphi, \phi\right)}
 \end{equation}
\end{Definition}

\begin{Proposition}\label{prop:muFM}
The function $\mu_{P^-}$ characterized in the Definition \ref{def:muP} is a fuzzy measure.
\end{Proposition}
\begin{proof}
To demonstrate this affirmation, we will show that the properties enunciated in the Definition \ref{def:fuzzMeasure} concerning fuzzy measures hold for $\mu_{P^-}$. 
\begin{itemize}
    \item $\mu_{P^-}\left(\emptyset\right)=0$. Trivial.
    \item $\mu_{P^-}\left(V\right)=1$. $\mu_{P^-}$ is 1-normalized by definition.
    \item Let $A,B \subseteq V$ such that $A\subseteq B$. Then, $\mu_{P^-}(A) \leq \mu_{P^-}(B)$. By definition, $JDJ_{pol}$ is a monotonic measure, so this property trivially holds. 
\end{itemize}
\end{proof}

\begin{Remark}\label{def:Pmenos}
Note that previous definition of the polarization fuzzy measure $\mu_{P^-}$ could be re-formulated as a summation concerning the different pairs of elements, i.e.,
\begin{equation}
    \mu_{P^-}\left(S\right)=\sum_{i,j\in S} P_{i,j}^{-}
\end{equation}
where 
\begin{equation}\label{eq:P-}
P_{i,j}^{-}= \frac{ \varphi \left( \phi ( \eta_{X_A}(i),\eta_{X_B}(j) ) ,\phi (\eta_{X_B}(i),\eta_{X_A}(j) ) \right)} {JDJ_{pol}\left(V,\eta_{X_A}, \eta_{X_B}, \varphi, \phi\right)}\end{equation}

Because of the properties of $\mu_{P^-}$,  $P^-$ is symmetric, non-negative, normalized, and its main diagonal is null. 

\end{Remark}

Because of the interpretation of the measure $JDJ_{pol}$, $P^-_{ij}$ represents the risk of conflict concerning the elements $i$ and $j$, so that $\mu_{P^-}$ represents the capacity of the elements to argue, to trigger conflict and arguments. Therefore, it is a recommended model to properly represent the discrepancy or distance between the individuals.

\begin{Example}\label{ex:calculaMuP}
In this example, we show the  calculation of $\mu_{P^-}$ for a given set $V$ with $4$ elements.
We consider the membership functions $\eta_{X_A}$ and $\eta_{X_B}$ defined in {Table} \ref{ta:membershipExample}.
We consider the functions $\varphi=\max$ and $\phi=product$. {Results are showed in Table} \ref{ta:valuesExample}.
\begin{specialtable}[H]
\caption{{Membership} degree of each element of $V$ to the poles $X_{A}$ and $X_{B}$.}
 
\setlength{\tabcolsep}{19mm}{\begin{tabular}{c c c }
\toprule
\textbf{Element} & \boldmath{${\eta_{X_A}}$} &  \boldmath{${\eta_{X_B}}$} \\ \midrule
1 &	1 &	0\\
2 &	0 &	1\\
3 &	1 &	0\\
4 &	0 &	1\\ \bottomrule
\end{tabular}}
\label{ta:membershipExample}
\end{specialtable}

Note that $JDJ_{pol}(V)=4$ is the amount of arguments among the $4$ elements, i.e., the capacity to trigger conflict. These conflicts come from the groups $\{1,2\}$ $\{1,4\}$, $\{2,3\}$, and $\{3,4\}$. 
\end{Example}

\end{paracol}
\nointerlineskip
\begin{specialtable}[H]
\widetable
\caption{{Values of the}  fuzzy measures $\mu_{P^-}$.}
\label{ta:valuesExample}
 
\setlength{\tabcolsep}{2.3mm}{\begin{tabular}{l|ccccccccccc}
\noalign{\hrule height 1pt}
${S}$ &  $\{1,2\}$ & $\{1,3\}$ & $\{1,4\}$ &  $\{2,3\}$ & $\{2,4\}$ & $\{3,4\}$ & $\{1,2,3\}$ & $\{1,2,4\}$ & $\{1,3,4\}$ &  $\{2,3,4\}$ & $\{1,2,3,4\}$\\ 
${\mu_{P^-}}\left(S\right)$ &	0.25 &	0 &	0.25 &	0.25 &	0 & 	0.25 &	0.5 &	0.5	& 0.5 & 0.5 &	1 \\ 
\noalign{\hrule height 1pt}
\end{tabular}}
\end{specialtable}
\begin{paracol}{2}
\switchcolumn

\begin{Remark}
We can define a measure obtained from the negation of the risk of polarization between two elements. That measure will have an opposite meaning than the capacity obtained from the $JDJ_{pol}$.
Let $N:[0,1]\rightarrow [0,1]$ denote a negation aggregator, and let us define

\begin{equation}\label{eq:noJDJ}
\widetilde{JDJ}\left(\{i,j\},\eta_{X_A},\eta_{X_B},\varphi,\phi\right)=N\left(\varphi(\phi(\eta_{X_A}(i), \eta_{X_B}(j)), \phi(\eta_{X_A}(j), \eta_{X_B}(i))\right)
\end{equation}

Then, we define the matrix $P^+$ as

\begin{equation}\label{eq:Pmas}
P^+_{ij}=\frac{\widetilde{JDJ}\left(\{i,j\},\eta_{X_A},\eta_{X_B},\varphi,\phi\right)}{\sum_{r,s\in V}\widetilde{JDJ}\left(\{r,s\},\eta_{X_A},\eta_{X_B},\varphi,\phi\right)}, \hspace{10mm} i,j\in V
\end{equation}
\end{Remark}

\begin{Definition}[Non-polarization fuzzy measure $\mu_{P^+}$]\label{def:muPmas}
Given a finite set $V$, a grouping function $\varphi$, a conjunction function $\phi$, a negation operator $N$, and two membership functions $\eta_{X_A},\eta_{X_B}:V\rightarrow [0,1]$, let $P^+$ be the matrix characterized in Equation \eqref{eq:Pmas}. Then, from matrix $P^+$, we can define a measure which represents the capacity of the elements of a set to peacefully dialogue without risk of Polarization:

\begin{equation}\label{eq:mupMas}
\mu_{P^+}\left(S\right)=\sum_{i,j \in S}P^+_{i,j}
\end{equation}
\end{Definition}

\begin{Remark}
Trivially, $\mu_{P^+}$ is a fuzzy measure.
\end{Remark}

\begin{Example}\label{ex:calculaMuPmas}
We recall Example \ref{ex:calculaMuP} in order to show the calculation  of $\mu_{P^+}$ for a given set $V$ with $4$ elements.
We consider the membership functions $\eta_{X_A}$ and $\eta_{X_B}$ defined in Table \ref{ta:toymuPmas}.
We consider the functions $\varphi=\max$, $\phi=product$, and $N(x)=1-x$. {Results are showed in Table} \ref{ta:resultsExample}.
\begin{specialtable}[H]
\caption{Membership degree of each element of $V$ to the poles $X_{A}$ and $X_{B}$.}
\label{ta:toymuPmas}
\setlength{\tabcolsep}{19mm}{\begin{tabular}{c c c }
\toprule
\textbf{Element} & \boldmath{${\eta_{X_A}}$} & \boldmath{${\eta_{X_B}}$} \\ \midrule
1 &	1 &	0\\
2 &	0 &	1\\
3 &	1 &	0\\
4 &	0 &	1\\ \bottomrule
\end{tabular}}
\end{specialtable}

Note that $\widetilde{JDJ}(V)=2$ is the amount of peaceful dialogues between the $4$ elements. These dialogues come from the groups $\{1,3\}$ and $\{2,4\}$.
\end{Example}

\clearpage
\end{paracol}
\nointerlineskip
\begin{specialtable}[H]
\widetable
\caption{V{alues }of the  fuzzy measures $\mu_{P^-}$ and $\mu_{P^+}$.}
\label{ta:resultsExample}
\widetable
\setlength{\tabcolsep}{2.3mm}{\begin{tabular}{l|ccccccccccc}
\noalign{\hrule height 1pt}
${S}$ &  $\{1,2\}$ & $\{1,3\}$ & $\{1,4\}$ &  $\{2,3\}$ & $\{2,4\}$ & $\{3,4\}$ & $\{1,2,3\}$ & $\{1,2,4\}$ & $\{1,3,4\}$ &  $\{2,3,4\}$ & $\{1,2,3,4\}$\\ 
${\mu_{P^-}}\left(S\right)$ &	0.25 &	0 &	0.25 &	0.25 &	0 & 	0.25 &	0.5 &	0.5	& 0.5 & 0.5 &	1 \\ 
${\mu_{P^+}}\left(S\right)$ &	0 &	0.5 &	0 &	0 &	0.5 &	0 &	0.5 &	0.5	& 0.5 & 0.5 &	1 \\\noalign{\hrule height 1pt}
\end{tabular}}
\end{specialtable}

\begin{paracol}{2}
\switchcolumn

Once we have defined two opposite models to represent the capacity of a set of elements to argue/dialogue, we define a new representation model:
 the polarization extended fuzzy graph.
 It combines the ability of a crisp graph to represent a set of elements connected to each other, with the representation of the synergies between these elements, regardless theirs connections. Therefore, from a crisp graph, two membership functions, and two aggregation operators, we can define a polarization extended fuzzy graph, a tool which sets light on the modeling of reality.
 
 \begin{Definition}[Polarization extended fuzzy graph]\label{def:EFGPolarized}
Let $G=(V,E)$ denote a crisp graph, whose nodes set is $V$ and whose edges set is $E$. Let $\eta_{X_A}$ and $\eta_{X_B}$ denote the membership functions of the elements of $V$ concerning the extreme poles $X_A$ and $X_B$. Let functions $\varphi: [0,1]^2\rightarrow [0,1]$ and $\phi: [0,1]^2\rightarrow [0,1]$ denote a grouping and a conjunction operator, respectively.  Let $\mu_{P^-}:2^V\rightarrow [0,1]$ according denote the fuzzy measure characterized in the Equation \eqref{eq:mup}. Then, the triplet $\widetilde{G}=(V,E,\mu_{P^-})$ is a polarization extended fuzzy graph.
\end{Definition}

Note that the representation ability of the polarization extended fuzzy graph goes far from the modeling provided by other tools as for example, a fuzzy graph. Let us show a toy example.

\begin{Example}\label{ex:polarizationEFG}
We consider the graph $G=(V,E)$ shown in the Figure. Let the membership functions $\left(\eta_{X_A}(1), \dots ,\eta_{X_A}(8)\right)=\left(1,0,0,1,1,1,0,0\right)$  \ and \ $\left(\eta_{X_B}(1), \dots ,\eta_{X_B}(8)\right)=\left(0,1,1,0,0,0,1,1\right)$ define the membership degree of each element of $V$  to the poles $X_A$ and $X_B$, respectively.  We consider $\varphi=\max$ and $\phi=product$.  In Figure \ref{fig:toyex:EFGPol}, we show a representation of the polarization extended fuzzy graph $\widetilde{G}=\left(V,E,\mu_{P^-}\right)$.

\begin{figure}[H]
\begin{subfigure}{0.39\linewidth}

 \includegraphics[scale=0.25]{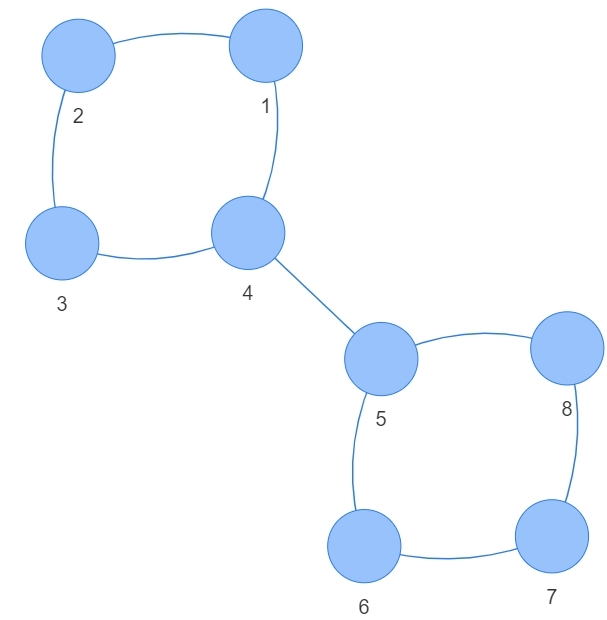}
\end{subfigure}
\begin{subfigure}{0.48\linewidth}\scalebox{1}{ 
$P^-=\frac{1}{32}\left( \begin{tabular}{c c c c c c c c}
0 & 1 & 1 & 0 & 0 & 0 & 1 & 1  \\
1 & 0 & 0 & 1 & 1 & 1 & 0 & 0  \\
1 & 0 & 0 & 1 & 1 & 1 & 0 & 0  \\
0 & 1 & 1 & 0 & 0 & 0 & 1 & 1  \\
0 & 1 & 1 & 0 & 0 & 0 & 1 & 1  \\
0 & 1 & 1 & 0 & 0 & 0 & 1 & 1  \\
1 & 0 & 0 & 1 & 1 & 1 & 0 & 0  \\
1 & 0 & 0 & 1 & 1 & 1 & 0 & 0  \\
\end{tabular}\right)$}
\end{subfigure}
    \caption{Polarization extended fuzzy graph $\widetilde{G}=\left(V,E,\mu_{P^-}\right)$.}
    \label{fig:toyex:EFGPol}
\end{figure}
\end{Example}

It may seem that the polarization extended fuzzy graph has a weak point related to the high complexity concerning the definition of the corresponding fuzzy measure $\mu_{P^-}$. Nevertheless, we will show some desirable properties of it, which facilitate the handling of $\widetilde{G}$. The most important is about the additivity, as it is shown below.

\begin{Proposition}\label{prop:2additive}
$\mu_{P^-}$ is a $2$-additive fuzzy measure.
\end{Proposition}

\begin{proof}
We base this demonstration on an asseveration found in \cite{gra2}, where Grabisch demonstrated that a fuzzy measure $\mu$ is 2-additive if and only if, for all $S\subseteq V$, it can be defined as a linear combination
$\mu(S)=\sum_{i=1}^n a_ix_i+\sum_{\{i,j\}\subset S}a_{ij}x_ix_j$, where $x_i=1 \textup{ if } i\in S \textup{, and } x_i=0 \textup{ otherwise}$.

$ $

For every $i\in V$, we define
$a_i=0$, and for every $i,j\in V$ such that $i\not= j$, we define $a_{ij}=P^-_{ij}$. 
Then, according to the Equation \eqref{eq:mup}, $\mu_{P^-}(S)=\sum_{i,j \in S}P^-_{ij}=\sum_{i,j \in V}P^-_{ij}x_ix_j=\sum_{i=1}^na_{i}+\sum_{i,j \in V}a_{ij}x_ix_j$.
\end{proof}

\begin{Proposition}\label{prop:muP:Comlin}
$\mu_{P^-}$ is closed for convex linear combinations:

\begin{equation*}
    \mu_{\left(\sum_{k=1}^m\alpha_k\widehat{P}^{-k}\right)}=\sum_{k=1}^m\alpha_k\mu_{\widehat{P}^{-k}}
\end{equation*}

where $\sum_{k=1}^m  \widehat{P}^{-k}=P^{-k}$ \ and \ $\sum_{k=1}^m\alpha_k=1$.
\end{Proposition}

\begin{proof}
According to Equation \eqref{eq:mup}, and assuming that $x_i=1 \textup{ if } i\in S \textup{, and } x_i=0 \textup{ otherwise}$, we have $$\mu_{\left(\sum_{k=1}^m\alpha_k\widehat{P}^{-k}\right)}=\sum_{i,j \in S} \left(\sum_{k=1}^m\alpha_k\widehat{P}_{ij}^{-k} \right)x_ix_j=\sum_{k=1}^m\sum_{i,j \in S}\alpha_k   \widehat{P}_{ij}^{-k} x_ix_j=$$ $$\sum_{k=1}^m\alpha_k\sum_{i,j \in S} \widehat{P}_{ij}^{-k}x_ix_j=\sum_{k=1}^m\alpha_k\mu_{\widehat{P}^{-k}}$$
\end{proof}

\begin{Remark}
As a particular case of the Proposition \ref{prop:muP:Comlin}, it holds that $\mu_P$ is fixed for the mean as follows:

\begin{equation*}
\frac{\mu_{P^{-1}}+\mu_{P^{-2}}}{2}=\mu_{\frac{P^{-1}+P^{-2}}{2}}
\end{equation*}
\end{Remark}

Note that all the points and properties enunciated with respect to $\mu_{P-}$ also apply to $\mu_{P^+}$. Then, $\mu_{P^+}$ is a $2$-additive fuzzy measure. Particularly, we emphasize in the definition of the non-polarization extended fuzzy graph $\widetilde{G}=\left(V,E,\mu_{P^+}\right)$, concerning a crisp graph and a non-polarization fuzzy measure.

\begin{Definition}[Non-polarization extended fuzzy graph]\label{def:nonPolEFG}
Let $G=(V,E)$ denote a crisp graph, whose nodes set is $V$ and whose edges set is $E$. Given a unidimensional variable $X$ with two extreme poles $X_A$ and $X_B$, let $\mu_{P^+}$ the non-polarization fuzzy measure characterized in the Definition  \ref{def:muPmas}. Then, the triplet $\widetilde{G}=(V,E,\mu_{P^+})$ is a non-polarization extended fuzzy graph.
\end{Definition}

\begin{Example}\label{ex:EFGnonPol}
We recall Example \ref{ex:polarizationEFG}, but in this case we focus on the measure $\mu_{P^+}$.
Therefore, we have the graph $G=(V,E)$ and  membership functions $\left(\eta_{X_A}(1), \dots ,\eta_{X_A}(8)\right)=\left(1,0,0,1,1,1,0,0\right)$  \ and \ $\left(\eta_{X_B}(1), \dots ,\eta_{X_B}(8)\right)=\left(0,1,1,0,0,0,1,1\right)$.  We consider $\varphi=\max$ and $\phi=product$, and $N(x)=1-x$. Then, the non-polarization extended fuzzy graph $\widetilde{G}=\left(V,E,\mu_{P^+}\right)$ is shown in Figure \ref{fig:toyex:EFGNoPol}, in which we show structure of the crisp graph and the matrix ${P^+}$ concerning $\mu_{P^+}$.

\begin{figure}[H]
\begin{subfigure}{0.39\linewidth}
 \includegraphics[scale=0.25]{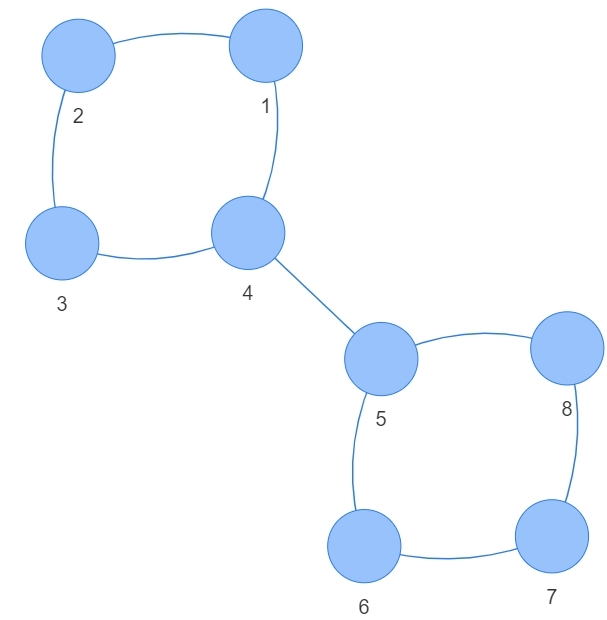}
\end{subfigure}
\begin{subfigure}{0.48\linewidth}\scalebox{1}{ 
$P^+=\frac{1}{24}\left(\begin{tabular}{cccccccc}
0 & 0 & 0 & 1 & 1 & 1 & 0 & 0  \\
0 & 0 & 1 & 0 & 0 & 0 & 1 & 1  \\
0 & 1 & 0 & 0 & 0 & 0 & 1 & 1  \\
1 & 0 & 0 & 0 & 1 & 1 & 0 & 0  \\
1 & 0 & 0 & 1 & 0 & 1 & 0 & 0  \\
1 & 0 & 0 & 1 & 1 & 0 & 0 & 0  \\
0 & 1 & 1 & 0 & 0 & 0 & 0 & 1  \\
0 & 1 & 1 & 0 & 0 & 0 & 1 & 0  \\
\end{tabular}\right)$}
\end{subfigure}
    \caption{Non-polarization extended fuzzy graph $\widetilde{G}=\left(V,E,\mu_{P^+}\right)$.}
    \label{fig:toyex:EFGNoPol}
\end{figure}
\end{Example}

Note that the measure $JDJ_{pol}$ quantifies  the distance or discrepancy between all pairs of elements ${i,j}$ of a given set of individuals $V$, i.e., the risk of polarization. Therefore, its negation $\widetilde{JDJ}_{pol}$ can be understood as the minimum risk of polarization for a given population or community. On this assumption, if we consider the non-polarization fuzzy measure $\mu_{P+}$, grouping  nodes according to this criterion allows us to build communities with minimum risk of conflict. These communities detected in the non-polarization extended fuzzy graph $\widetilde{G}=\left(V,E,\mu_{P^+}\right)$ present a structure that better fixes reality than those communities built only by the relations between nodes.


As the aim of this paper is related to community detection problems, hereafter we will focus on the non-polarization fuzzy measure $\mu_{P^+}$ and the corresponding matrix $P^+$. Both tools allow us to manage the synergies between the nodes. To simplify the notation, we  consider $\mu_P = \mu_{P^+}$ and $P = P^+$.

\section{A Parametric Approach to Community Detection Problem Based on Polarization Measures and Weighted Mean}\label{sec:CDPPolari}
Many complex networks show a modular structure so that the individuals are organized into modules with dense internal connections. Numerous examples can be found:  in the field of social networks, groups of related users according to their interests or background; in any citation network, groups of connected papers concerning one particular issue; in a recommendation network, set of similar services or offers; and in metabolic networks, connected biochemical pathways \cite{biological,guimera,selfOrganization}. Due to the increasingly demand for all these real-life applications among many others, having a consistent community structure helps to understand the main characteristics, functions, and topology of these systems. So that, a good understanding of the community structure hidden in a complex network may be helpful for  better analysis and exploitation of the data in an effective way \cite{discreteBack,penalized}. 

Complex networks are usually represented by graphs. One of their most popular applications is devoted to the resolution of community detection problems, whose main goal is to find a \textit{good} partition of a given network. A partition of a graph $G=(V,E)$ is a decomposition of the set of nodes $V$ into subgroups known as communities or clusters whose composition depends on the similarity between the objects considered, i.e., a division of the set of nodes into groups that are densely intra-connected, whereas sparsely connected with the rest of the graph \cite{fortunato,guptaParallel,vitalli}.

Classical algorithms proposed in this field are based on topological information and on the structure of the network considered.
Nevertheless, it is undeniable that in the process of modeling the reality by means of a network for subsequent groups search, there is lot of knowledge and information that are not considered in the grouping process.
Several authors agree on the importance of adding some additional information to the structure represented by a graph to enrich the communities detected \cite{covec,consistencyRiolo,de2018combined,centralityPower}. In this work,   the  problem addressed by Gutiérrez et al. \cite{infus,ampliinfus,escim19,multipleBip,wcci2020,ipmu2020,flins2020} about the detection of communities in extended fuzzy graphs is particularly interesting. They proposed a methodology to analyze independently the structural information of the graph and the knowledge represented by the fuzzy measure when grouping the nodes.

We approach the community detection problem based on fuzzy measures including this information about the relations among the individuals emanating from that knowledge about the respective positions in any attitudinal axis. These relations will be considered in terms of Polarization measures built from the $JDJ_{pol}$ measure. Then, the base of the problem here addressed is a non-polarization extended fuzzy graph  $\widetilde{G}=(G=(V,E),\mu_P)$. 
Note the increment of cohesiveness procured in the groups by considering additional information independent of the topology. Note that we consider the non-polarization fuzzy measure $\mu_P=\mu_{P^+}$ instead of the polarization fuzzy measure $\mu_{P^-}$ in order to fix an scenario in which all the components of $\widetilde{G}$, $A$ and $\mu_P$, have the same somewhat ``positive'' \ nature.

    To face this problem, we work inspired by the idea developed in the Additional Louvain algorithm (see in \cite{ampliinfus}), based on the Louvain algorithm \cite{blondel}. The key point is to distinguish two different roles within the input parameters: one of them, to establish the neighbor relations, and the other, to calculate the variation of the modularity. The first role will be played by the adjacency matrix of the graph, $A$, so that only those nodes that are connected in $G$ can be in the same group. On the other hand, we suggest to consider a combination of the two components of the non-polarization extended fuzzy graph $\widetilde{G}$ as basis to calculate the variation of modularity, in order to incorporate the additional information. 
Then, having a crisp graph $G$, the two membership functions $\eta_{X_A}$ and $\eta_{X_B}$, the operators $\varphi$ (grouping), and $\phi$ (overlapping) and considering the negation function $N$, or what is the same, a non-polarization extended fuzzy graph $\widetilde{G}=\left(V,E,\mu_P\right)$, we propose a new methodology that is summarized as follows.

\begin{enumerate}
    \item Obtain the non-polarization fuzzy measure $\mu_P$ related to the set $V$ from the parameters $\eta_{X_A}$, $\eta_{X_B}$, $\varphi$, $\phi$ and $N$, according to Equation \eqref{eq:mupMas}.
    \item Summarize $\mu_P$ into a matrix, $F$.
    \item Define the matrix $M=\theta(A,F)$, where $\theta:\Pi(n)^2\rightarrow \Pi(n)$ is a matrix aggregator used to combine two matrices into a single one.
    \item Apply the Louvain algorithm by distinguishing the role of the matrices $A$ and $M$: $A$ is used to find the neighbor relations, $M$ is used to calculate the variation of modularity.
\end{enumerate}

\begin{Remark}
In this proposal, we  suggest the use of a matrix aggregator $\theta$. Nevertheless, any other operator could be applied instead.
\end{Remark}
The definition of the matrix $F$, as an aggregation of the non-polarization fuzzy measure $\mu_P$, should be closely related to the problem addressed. We suggest a particular characterization of it, based on the calculation of the weighted graph associated with $\mu_P$.
This matrix is a highly recommended tool for fuzzy measures manipulation and visualization, which summarizes the knowledge about the capacity of the elements into $n^2$ data set. The definition of this graph is based on the Shapley value \cite{shapley}, particularly in its characterization related to fuzzy measures \cite{uncertainty}.

\begin{Definition}[Weighted graph associated with a fuzzy measure ${G_{\mu}}$ \cite{infus,ampliinfus}]\label{def:associatedGraph}
Let $\mu:2^V\rightarrow [0,1]$ denote a fuzzy measure defined over the finite set $V$, and let $\xi:[-1,1]^2\rightarrow [0,1]$ denote a bivariate aggregation operator. We consider $Sh_i(\mu)$, the Shapley value of the individual $i\in V$ in coalition with all the elements of $V$ regarding their relation in $\mu$; analogously, $Sh_i^j(\mu)$ denotes the Shapley value of the individual $i$ in coalition with all the elements of $V\backslash\{j\}$, regarding $\mu$. Then, the weighted graph associated with the fuzzy measure $\mu$, denoted by $G_{\mu}$, is that whose adjacency is represented by the matrix $ F $, where

\begin{equation}\label{eq:associatedGraph}
     F _{ij}=\xi \left( Sh_i(\mu)-Sh_i^j(\mu), Sh_j(\mu)-Sh_j^i(\mu)\right) 
\end{equation}
\end{Definition}

In our specific proposal of the method to find communities in a non-polarization extended fuzzy graph, we suggest summarizing the non-polarization fuzzy measure $\mu_P$ into the matrix $F$, with adjacency of its associated weighted graph
. To formally establish this method, let us define it as an algorithm, named Polarization Louvain, whose pseudocode can be found in Algorithm \ref{alg:AdditionalLouvainPol}.

\begin{algorithm}[H]
	\begin{algorithmic}[1]
	\STATE\textbf{Input}: $\left(A, \  \eta_{X_A}, \eta_{X_B}, \varphi, \phi \right)$;
		\vspace{1mm}
		\STATE\textbf{Output}: $P$;
		\vspace{1mm}
		\STATE\textbf{Preliminary}
		\vspace{1mm}
		\STATE  $\mu_P \leftarrow \left(\eta_{X_A}, \eta_{X_B}, \varphi, \phi, N\right)$;
		\vspace{1mm}
		\STATE  $F _{ij}=\xi \left( Sh_i(\mu_P)-Sh_i^j(\mu_P), Sh_j(\mu_P)-Sh_j^i(\mu_P)\right)$, \ for all $i,j\in V$;
		\vspace{1mm}
		\STATE $M \leftarrow\theta(A, F)$;
		\vspace{1mm}
	\STATE $C_i \leftarrow \{i\}$, \  $\forall i \in V$ \ \ (each node is an isolated community);
			\vspace{1mm}
	\STATE $P \leftarrow \left(1, 2 \ldots, n\right)$ \ \ \ \ (initial partition);
			\vspace{1mm}
			\STATE\textbf{end Preliminary}
			\vspace{1mm}
\STATE\textbf{Phase 1}
			\vspace{1mm}
\STATE $\left(o^1, \dots, o^i,\ldots, o^n\right) \leftarrow perm(V)$; 
			\vspace{1mm}
	\STATE $stop \leftarrow 0$;
			\vspace{1mm}
	\WHILE{$(stop==0)$}
	\vspace{1mm}
	\STATE{$stop \leftarrow 1$}
			\vspace{1mm}
	    	\FOR{$(i=1)$ \TO $(n)$}
	\STATE{$H\left(o^i\right)\leftarrow\left(e_1,\dots, e_h\right)$ \ \ (find  all the neighbours of $o^i$ in $A$);}		
			\vspace{1mm}
	\FOR{$(j=1)$ \TO $(h)$}
			\vspace{1mm}
	\STATE{ Compute $\Delta Q_{o^i}(e_j)$ in  ${M}$;}
			\vspace{1mm}
	\ENDFOR
			\vspace{1mm}
	\STATE{${\displaystyle j^* \leftarrow \Big\{e_{\ell} \ | \ \Delta Q_{o^i}(j^*)=\max_{\ell\in\{1 \dots , h\}}\big\{\Delta Q_{o^i}(e_{\ell})\big\}\Big\}}$;}
			\vspace{1mm}
	\IF{$(\Delta Q_{o^i}(j^*)>0)$}
			\vspace{1mm}
	\STATE $C_{P\left(o^i\right)} \leftarrow C_{P\left(o^i\right)}\backslash{\{o^i\}}$; 
		\vspace{1mm}
	\STATE $C_{P\left(j^*\right)} \leftarrow C_{P\left(j^*\right)} \cup \{o^i\}$; 
		\vspace{1mm}
	\STATE $P\left(o^i\right) \leftarrow P\left(j^*\right)$;	
		\vspace{1mm}
	\STATE $stop \leftarrow 0$;
			\vspace{1mm}
	\ENDIF
			\vspace{1mm}
	\ENDFOR
				\vspace{1mm}
\ENDWHILE
				\vspace{1mm}
	\STATE\textbf{end Phase 1}
				\vspace{1mm}
	\STATE\textbf{Phase 2}
	\vspace{1mm}
\STATE{Aggregate $A^*$  from  $A$  (nodes of $A^*$ are the communities found in Phase 1);}
\vspace{1mm}
\STATE{Aggregate $M^*$  from  $M$  (nodes of $M^*$ are the communities found in Phase 1);}
\vspace{1mm}
\IF {$(A^* \neq A)$}
	\STATE $A \leftarrow A^*$;
	\vspace{1mm}
	\STATE $M \leftarrow M^*$;
	\vspace{1mm}
	\STATE Compute \textbf{Phase 1} and \textbf{Phase 2};
	\vspace{1mm}
	\ENDIF
	\vspace{1mm}
\STATE\textbf{end Phase 2}
			\vspace{1mm}
	\STATE\textbf{return}$\left(P\right)$;	\end{algorithmic}\caption{\textit{Polarization Louvain}}
	\label{alg:AdditionalLouvainPol}
\end{algorithm}

 $ $

The key point to approach a clustering process in a  non-polarization extended fuzzy graph $\widetilde{G}=(V,E,\mu_P)$  is the calculation of the weighted graph associated with $\mu_P$. The calculation of the Shapley value on which it is based is a process that usually reaches exponential complexity. Nevertheless, we will show that this problem does not apply when considering $\mu_P$, for which we have demonstrated in Proposition \ref{prop:2additive} that it is a $2$-additive fuzzy measure.

\begin{Proposition}\label{prop:shapleyP} Let $\mu_P$ denote the non-polarization fuzzy measure related to set $V$ obtained from the membership functions $\eta_{X_A}$ and $\eta_{X_B}$, the aggregation operators $\varphi$ and $\phi$, and the negation operator $N$, according to Definition \ref{def:muPmas}, so that  $P$ is the matrix obtained from these parameters according to Equation \eqref{eq:Pmas}. The following holds for $\mu_P$.
		\begin{enumerate}
		\item $$Sh_i(\mu_P)=\sum_{k\in V}	P_{ik}$$
		
		\item $$Sh_i^j(\mu_P)=\sum_{k\in V\backslash\{j\}}P_{ik}=\left(\sum_{\\k\in V}P_{ik}\right) - P_{ij} $$
	\end{enumerate}
\end{Proposition}

\begin{proof}
We prove the point $1$, so that the demonstration of $2$ is analogous. 

It is based on an alternative characterization of the Shapley value \cite{imprSh,poliSh} in which, $\forall i \in V$, the corresponding Shapley index can be calculated as the average of the marginal contributions in all the permutations of the original set $V$, i.e.,
$$Sh_i=\frac{1}{n!}\sum_{o\in\pi(n)}[\mu\left(pred(i)+\{i\}\right)-\mu\left(pred(i)\right)]$$ where $pred(i)$ denotes the set of predecessors of $i$ in the order $o$ and $\pi(n)$ denotes the set of all the possible permutations of a set with $n$ elements. 

According to Equation \eqref{eq:mupMas}, $$\mu\left(pred(i)+\{i\}\right)=\sum_{\begin{subarray}{1}k=1  \end{subarray}}^n\sum_{\begin{subarray}{1}j=1\end{subarray}}^nP_{jk}x_jx_k+\sum_{k=1}^n\left(P_{ik}+P_{ki}\right)x_k$$ $$\mu\left(pred(i)\right)=\sum_{\begin{subarray}{1}k=1  \end{subarray}}^n\sum_{\begin{subarray}{1}j=1 \end{subarray}}^nP_{jk}x_jx_k$$ being $x_j=1 \textup{ if } j \in pred(i), \textup{ and } x_j=0 \textup{ otherwise }$.
    
    $ $
    
So that, 

\begin{adjustwidth}{-4.6cm}{0cm}
$$Sh_i=\frac{1}{n!}\sum_{o\in\pi(n)}\left(\sum_{\begin{subarray}{1}k=1  \\\end{subarray}}^n\sum_{\begin{subarray}{1}j=1  \end{subarray}}^nP_{jk}x_jx_k+\sum_{k=1}^n\left(P_{ik}+P_{ki}\right)x_k\right)-\left(\sum_{\begin{subarray}{1}k=1  \end{subarray}}^n\sum_{\begin{subarray}{1}j=1 \end{subarray}}^nP_{jk}x_jx_k\right)= \frac{1}{n!}\sum_{o\in\pi(n)}\sum_{k=1}^n\left(P_{ik}+P_{ki}\right)x_k$$
\end{adjustwidth}

For a half of the orders $o\in \pi(n)$, it is true that $k\in pred(i)$, so, for a half of the values of the previous summation,  $x_k=1$. Therefore, 

$$\frac{1}{n!}\sum_{o\in\pi(n)}\sum_{k=1}^n\left(P_{ik}+P_{ki}\right)x_k=\frac{1}{2}\sum_{k=1}^n(P_{ik}+P_{ki})$$

By definition, $P$ is symmetric, and its main diagonal is null. Therefore, $P_{ik}=P_{ki}$ and $P_{ii}=0$. Then,

$$\frac{1}{2}\sum_{k=1}^n(P_{ik}+P_{ki})=\frac{1}{2}\sum_{k=1}^n2P_{ik}=\sum_{k=1}^nP_{ik}=\sum_{\begin{subarray}{1} k \in V\end{subarray}}P_{ik}$$
\end{proof}

As a consequence of the Proposition \ref{prop:shapleyP}, the following result holds for $\mu_P$.

\begin{Proposition}\label{prop:ShP:afinidad}
Let $\mu_P$ denote the non-polarization fuzzy measure related to set $V$ obtained  from the membership functions $\eta_{X_A}$ and $\eta_{X_B}$, the aggregation operators $\varphi$ and $\phi$, and the negation operator $N$, according to Definition \ref{def:muPmas}, so that  $P$ is the matrix obtained from these parameters according to Equation \eqref{eq:Pmas}.
Let $i,j\in V$ denote two individuals. Then, the following applies.
	 \begin{equation*}
	 1. \ \ Sh_i(\mu_P)-Sh_i^j(\mu_P)=P_{ij}
	 \end{equation*}
	 	 \begin{equation*}
	 2. \ \ Sh_j(\mu_P)-Sh_j^i(\mu_P)=P_{ji}
	 \end{equation*}
\end{Proposition}

\begin{proof}
We prove the point $1$, so that the demonstration of $2$ is analogous.

As it is demonstrated in the  Proposition \ref{prop:shapleyP}, $$1. \ \ Sh_i(\mu_P)-Sh_i^j(\mu_P) = \sum_{\substack{k\in V}}	P_{ik}-\left(\sum_{\substack{k\in V}}P_{ik} - P_{ij}\right)=\sum_{\substack{k\in V}}	P_{ik}-\sum_{\substack{k\in V}}P_{ik} + P_{ij}=P_{ij}$$
\end{proof}	 

At this point, it is trivial to represent the closeness between different pairs of elements according to their attitude concerning a particular issue. Then, as $\mu_P$ is the corresponding non-polarization fuzzy measure based on the Polarization measure $JDJ_{pol}$, we assume that the closeness between two individuals concerning its attitude about one issue can be represented by the weighted graph associated with $\mu_P$, i.e., with the corresponding adjacency matrix of $G_{\mu_P}$, calculated as
$$F_{ij}= \xi \left(Sh_i(\mu_P)-Sh_i^j(\mu_P), Sh_j(\mu_P)-Sh_j^i(\mu_P)\right)=\xi\left(P_{ij},P_{ji}\right)$$

\begin{Remark}
Note that because of $P$'s symmetry, if the chosen aggregation operator $\phi$  is of the type $max$, $min$, $average$ among others, then $ F _{ij}=P_{ij}$, $\forall i,j\in V$.
\end{Remark}

So far, we have summarized all the knowledge modeled by the non-polarization extended fuzzy graph $\widetilde{G}=(V,E,\mu_P)$ into two independent matrices, $A$ and $F$. This process/tools could be applied in many fields, as, for example, problems about centrality, link prediction or propagation.

It is crucial to be clear about the interpretation of the matrices $A$ and $F$ (or  $P$ in any case). On the one hand, $A$ represents the direct connections between the elements of $V$; it is well accepted that nodes tightly-knit connected should be connected, so it can be seen that the connections shown in $A$ are ``positive''. On the other hand, we have already mentioned that, because of the characterization of $\mu_P$ (and thus of $P$/$F$), it is related to the synergies or closeness between the elements. Then, we can conclude that both matrices $A$ and $F$ have ``positive'' meanings, so that nodes for which both matrices (or even one of them if it is fair enough) define high values, should be together. 

Let us illustrate the performance of the Polarization Louvain method with a toy example.
In this case, we combine the matrices $A$ and $F$ by means of a linear combination ($\theta(A,F)=\gamma A + (1-\gamma)F$). In our opinion, it is an smart way to assign a weight or importance to each component of the $\widetilde{G}$. Note that, when $\gamma=1$, the additional information is not considered. In this case,  both the search of neighbor relations and the modularity variation are calculated over the matrix $A$, so that the Polarization Louvain algorithm is exactly the same than the Louvain algorithm.    

\begin{Example}\label{ex:CDPol}
Let us consider the graph $G=(V,E)$ whose adjacency matrix is $A$, and let us assume some knowledge about the position of the elements of $V$ in any attitudinal axis modeled, in the sense that we know the membership degree of all the individual in $V$ to the poles $X_A$ and $X_B$, represented by the membership functions $\eta_{X_A}$ and $\eta_{X_B}$, respectively. { These values are showed in Table} \ref{ta:example7}. From this knowledge, and considering the operators $\varphi=\max$ and $\phi=prod$ and $N(x)=1-x$, we define the fuzzy measure $\mu_P$, and therefore the matrix $P$ (Equation \eqref{eq:mupMas}).

\begin{specialtable}[H]
\caption{{Membership degree of each node} of $V$ to the poles $X_{A}$ and $X_{B}$.}
\label{ta:toyPol}
\setlength{\tabcolsep}{18.6mm}{\begin{tabular}{c c c }
\toprule
\textbf{Vertex} & \boldmath{${\eta_{X_A}}$} & \boldmath{${\eta_{X_B}}$} \\ \midrule
1 &	0.022 &	0.878\\
2 &	0.756 &	0.144\\
3 &	0.751 &	0.099\\
4 &	0.5 &	    0.5\\
5 &	0.001 &	0.989\\
6 &	0.102 &	0.888\\
7 &	0.889 &	0.112\\ \bottomrule
\end{tabular}}\label{ta:example7}
\end{specialtable}

Due to the properties of  $P$, we have that $F=P$, so the characterization of $\mu_P$ is straightforward.

At this point, we have the non-polarization extended fuzzy graph $\widetilde{G}=(V,E,\mu_P)$, so that, being $\gamma\in[0,1]$ a balancing factor, and considering the aggregation function $\theta(A,F)=\gamma A + (1-\gamma)F=\gamma A + (1-\gamma)P$,  we apply the Polarization Louvain algorithm, {being $A$ and $P$ the matrices showed in Figure} \ref{fig:toyex:matricesPOL}. In Figure \ref{fig:partitionsToyExamplesPOL}, we show the partitions obtained for several values of $\gamma$. Note how the way in which the nodes are organized changes depending on the importance assigned to the information represented by $P$ (i.e., $F$) about the closeness between the nodes.
\end{Example}
\end{paracol}
\nointerlineskip
\begin{figure}[H]
\widefigure
\begin{subfigure}{0.39\linewidth}\scalebox{0.9}{ 
$A=\left( \begin{tabular}{ccccccc}
0 & 1 & 1 & 0 & 1 & 0 & 1\\
1 & 0 & 1 & 0 & 0 & 0 & 0\\
1 & 1 & 0 & 1 & 1 & 0 & 1\\
0 & 0 & 1 & 0 & 1 & 0 & 0\\
1 & 0 & 1 & 1 & 0 & 1 & 1\\
0 & 0 & 0 & 0 & 1 & 0 & 1\\
1 & 0 & 1 & 0 & 1 & 1 & 0 \\
\end{tabular}\right)$}
\end{subfigure}
\begin{subfigure}{0.48\linewidth}\scalebox{0.9}{ 
$ P =\frac{1}{22.574}\left( \begin{tabular}{ccccccc}
0.000 & 0.336 & 0.341 & 0.561 & 0.978 & 0.910 & 0.219\\
0.336 & 0.000 & 0.892 & 0.622 & 0.252 & 0.329 & 0.872\\
0.341 & 0.892 & 0.000 & 0.625 & 0.257 & 0.333 & 0.912\\
0.561 & 0.622 & 0.625 & 0.000 & 0.506 & 0.556 & 0.556\\
0.978 & 0.252 & 0.257 & 0.506 & 0.000 & 0.899 & 0.121\\
0.910 & 0.329 & 0.333 & 0.556 & 0.899 & 0.000 & 0.211\\
0.219 & 0.872 & 0.912 & 0.556 & 0.121 & 0.211 & 0.000\\
\end{tabular}\right)$}
\end{subfigure}
    \caption{{Matrices} $A$ and $P$.} 
    \label{fig:toyex:matricesPOL}
\end{figure}
\begin{paracol}{2}
\switchcolumn

\begin{figure}[H]
\begin{subfigure}{0.3\linewidth}

    \includegraphics[scale=0.31]{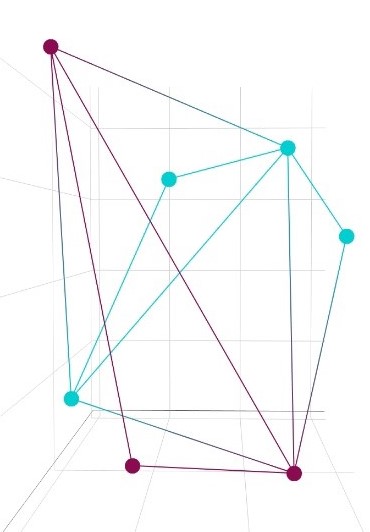}
    \caption{$\gamma=1$ (Louvain algorithm)}\label{fig:alpha1POL}
    \end{subfigure}
\begin{subfigure}{0.3\linewidth}

    \includegraphics[scale=0.31]{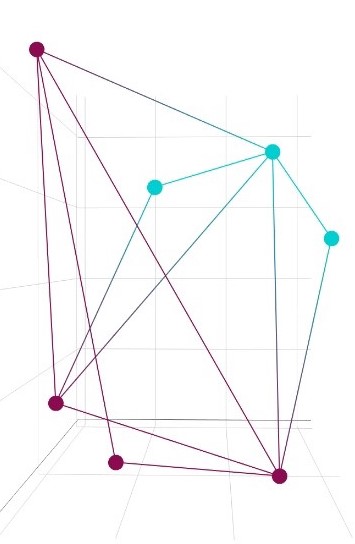}
    \label{fig:alpha5POL}
    \caption{$\gamma=0.5$}\end{subfigure}
\begin{subfigure}{0.3\linewidth}

    \includegraphics[scale=0.31]{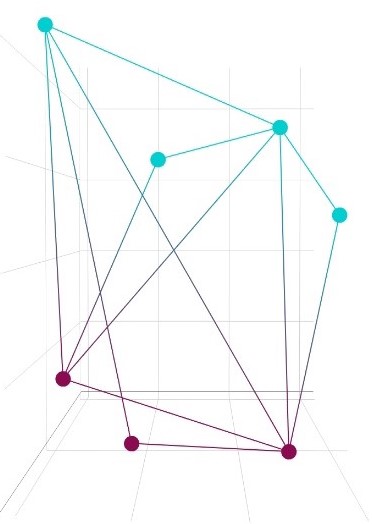}
    \label{fig:alpha0POL}\caption{$\gamma=0$}
    \end{subfigure}   
\caption{Partitions of  $\widetilde{G}=(V,E,\mu)$ obtained with the Polarization Louvain algorithm.}\label{fig:partitionsToyExamplesPOL}\end{figure}

\section{A Real Case: The Impact of the COVID-19 Pandemic in the Organization of the People }\label{sec:realcase}
\subsection{Experiment Design: Sources and Methodology}\label{sec:experiment}
We briefly explain the case of study in the following. The nodes and theirs relations considered in this work have been obtained from the social network Twitter, particularly from some posts recorded along the state of alarm imposed by the central government in Spain (from March 16th, 2020 until June 29th, 2020). All data downloaded relate to the COVID-19 pandemic and the political situation in that country, concerning the management of the sanitary situation by the Spanish government. Each element of that  data set  represents an influential and verified account.

It is well known by popular knowledge that Twitter is one of the trendiest online social networks, where millions of users debate about any social or political topic, among many others. For this research, we have used the retweet (RT) network. A RT is a post derived from the action of any user to replicate a given tweet or message of another user to spread that content with his/her followers. In the literature, the RT network has been commonly used  as a directed network \cite{robles2019polarizacion}, so that if the original tweet is written by an user $i$ and then $j$ RT it, then there is a connection with directionality, which represents the action of each user.  Nevertheless, in our case we understand it as an non-directed network in this sense: it is not of our matter to know the directionality of a connection (who posts the tweet and who RT it), but rather to focus on the content. In this vein, once a given user $j$ RT a tweet of $i$, what is important to us is the intention of $j$ to transmit and spread that content. In broader terms, we can assume that $j$ agrees with the content and spreads the word to make the tweet visible and influential over the people. Our aim is to know the user's political attitude towards the Spanish government measuring their attitude reflected on the tweets. Therefore, no directionality is needed.

All data were downloaded from Twitter, using its API by R-Studio, with the package ``rtweet'' \cite{rtweet-package} in 5 rounds along the state of alarm in Spain.

$ $

\begin{tabular}{l c c c}
    $\triangleright$ 1st round: & ``2020-03-16'' &  -- & ``2020-03-23''. \\
    $\triangleright$  2nd round: & ``2020-04-06'' &  -- & ``2020-04-21''.\\
    $\triangleright$ 3rd round: & ``2020-05-07'' & -- & ``2020-05-22''. \\
    $\triangleright$ 4th round: & ``2020-06-03'' & -- & ``2020-06-15''.\\
    $\triangleright$ 5th round: & ``2020-06-14'' & -- & ``2020-06-29''.\\
 \end{tabular} 

$ $

The criteria we used to download the tweets were related to the considerations of those keywords which are mainly composed by the main political parties in Spain as well as their leaders:
\\\\
\textit{psoe OR pp OR vox OR ciudadanos OR gobierno OR podemos OR españa OR sanchezcastejon OR vox\_es OR pabloiglesias OR pablocasado\_ OR santi\_abascal OR inesarrimadas OR CiudadanosCs OR populares OR estadodealarma}
\\

After the downloading phase, we obtained 4.895,747 tweets, {about which we know, among other points, who posted it and who retweeted it}. Then, manual encoding was applied in a sample of those tweets to fix the points:
\begin{enumerate}
\item[$(1)$] To detect and filter all those tweets included on our database which do not correspond to our goals (Feature: {\textbf{TOPIC}}). 
\item[$(2)$] To encode each tweet as  (a) detractor, (b) neutral, or (c) supporter of the Spanish govern  ment (Feature: \textbf{POSITION}). 
\end{enumerate}

The manual encoding was applied to a random sample of $1500$ tweets for each round mentioned above. {To carry on with it, we analyzed the subject matter of the tweet. As we are considering extreme positions, from our personal knowledge about the political position, it is not complicated to decided whether the message of a tweet is in favor of the government, against it, or neutral.} Note the importance of encoding by rounds due to the dynamic nature of debates on online social networks, in which words or events can change over time despite being debating about one specific topic. Once the data were encoded, we applied text classification with  machine learning algorithms in order to tackle the full content of our database. 

According to the work in \cite{wang2006optimal}, \textit{Linear Support Vector Machines} are recognized to be one of the best machine learning algorithms for text classification. So that, after the \textit{tokenization} phase and the removal of \textit{stopwords}, we converted our text into a \textit{tf-idf} matrix. This type of matrices presents all the different words which appeared on the corpus on the columns, and the strings (tweets in our case) on the rows. The simplest \textit{dfm} matrix is an occurrence matrix with 0 if a given word does not appear on the tweet and 1 if it does. However, \textit{tf-idf} matrices show values as a result of the product of a term frequency and inverse document frequency for each word of a tweet. So that, as it is a classification problem with the classes \textit{detractor} (pole $X_A$) and \textit{supporter} (pole $X_B$), the classifier was trained and applied for the feature ``TOPIC'' and then for the ``POSITION''. The results obtained with that process of text classification are showed in the Table \ref{tab:table_SVM}. Note that the final scores recorded for ``POSITION'' are the two probabilities for being a ``detractor'' or ``supporter'' tweet towards the Spanish government. In this case, the ``neutral'' category is omitted in order to get a variable with two poles, assuming that probabilities close to $0.5$ correspond to the ``neutral'' category. {To carry on with this phase, we used the R-package \textit{e1071}, where all these methods of text classification are implemented \cite{Re1071}.} 

In Table \ref{tab:table_SVM}, we show an analysis in which the following indices have been considered:  \textit{PRECISION}, \textit{RECALL}, \textit{KAPPA}, \textit{F-SCORE} , and \textit{AUC} \cite{AUC,joachim}.
\begin{specialtable}[H]
 
    \caption{Linear support vector machine (SVM) performance for features ``TOPIC'' and ``POSITION''.}
    \label{tab:table_SVM}
\setlength{\tabcolsep}{3.8mm}{\begin{tabular}{c c c c c c c}
    \toprule
      \textbf{Round} & \textbf{Feature} & \textbf{Precision} & \textbf{Recall} & \textbf{Kappa} & \textbf{F-Score} & \textbf{AUC}\\
      \midrule
        1&  TOPIC & 0.8017 & 0.9322 & 0.3670 & 0.8620 & 0.6583\\
        2&	TOPIC&	0.8167&	0.5476&	0.5077&	0.6556&	0.7344\\
        3&	TOPIC	&0.8267	&0.7027	&0.6187&	0.7596	&0.8010\\
        4&	TOPIC&	0.7867&	0.7090&	0.564&	0.7457&	0.7791\\
        5&	TOPIC&	0.7659&	0.8758&	0.5216&	0.8171&	0.7567\\
        1&	POSITION&	0.8492&	0.9854&	0.4816&	0.9122&	0.6950\\
        2&	POSITION&	0.8960&	0.9619&	0.7761&	0.9277&	0.8780\\
        3&	POSITION&	0.8392&	0.8488&	0.6675&	0.8439&	0.8366\\
        4&	POSITION&	0.9133&	0.9048	&0.8225&	0.9090&	0.9121\\
        5&	POSITION&	0.8318&	0.8600	&0.6638	&0.8456&	0.8335\\
        \bottomrule
\end{tabular}}
 
\end{specialtable}

Finally, the database derived from the \textit{SVM} classifier is integrated by  1,208,631  tweets {which have been posted or RT by}  469,616  different users. To aim for those \textit{influencers} and verified accounts, we filtered by the following:
\begin{enumerate}
\item[$(a)$] Tweets with high repercussion on Twitter, considering accounts whose tweets with RT count are placed above the $50$ percentile ($n \geq 317$). {The information about the count of RT is provided by the API.}
\item[$(b)$] Verified accounts. {This information is provided by the Twitter API by means of a logical variable which indicates if a given tweet has been posted/RT by a verified account or not.}
\item[$(c)$] Accounts with high number of followers, considering accounts whose number of followers is placed above the $50$ percentile ($n \geq $21,779). {The information about the number of followers is provided by the API.}
\end{enumerate}

In this manner, $406$ users left, mainly politicians, party accounts, and journalists. Then, to get a closed network of users, we matched those accounts that both write or RT any tweet among those $406$ accounts. So that in our final data base, $295$ users are considered, from whose posts,  $657$ interactions are derived. Note that these interactions may concern users who are not among the $295$ considered, but who have RT some of theirs posts; so that we have a total amount of $454$ different users and $657$ interactions. {Each user will be represented by a node, and each interaction by a non-directed and non-weighted edge.}  From this information, we build a network $G=(V,E)$, so that the each of these $454$ accounts is represented by a node of the set $V$, and the links represents the edges of $E$. Let us remark that we take into account if it comes the case in which two users interact several times (by means of RT of different tweets), i.e., we work with a weighted graph, so that weight $w_{ij}$ of the corresponding edge represents how many time have interacted the users $i$ and $j$.

Note that the objects to be classified were not users but tweets, so, for each user, we computed  the average score for his/her tweets of being ``detractors'' and ``supporters'' (not only considering the original posts, but also the RT). At this end, we finally got, for the $295$ users, two specific values of probability for being a ``detractor'' and  for being a ``supporter'' toward the Spanish government, {provided by the SMV method}. These distances to the support vector machines for each class of the SMV will be used as membership degree values for each user to compute the $JDJ_{pol}$ Polarization measure,  which sets the basis of the non-polarization measure $\mu_P$ (Definition \ref{def:muPmas}), which is part of the non-polarization extended fuzzy graph $\widetilde{G}=\left(V,E,\mu_P\right)$ over which we apply the community detection problem.

For a  better understanding of the results, it is important to provide a proper visualization of the network, which comprises a complex process \cite{networkVisualization}. Having a proper organization and a good representation of the network itself is fundamental for a better understanding and exploitation of the inherent data. To accomplish that, we have used the R package ``visNetwork'' \cite{visNetwork}.

\subsection{Results}\label{sec:results}
In this section, we show the computational results obtained when applying the Polarization Louvain algorithm to the data set obtained from Twitter as explained in \mbox{Section \ref{sec:experiment}}. To carry on with it, we build a graph from that obtained data set and a fuzzy measure which represents the capacity of the elements to trigger conflict. Originally, this graph had $454$ nodes and $657$ weighted edges (the list of the interactions between users from which we define the set of edges can be found in {GitHub}~({{\url{https://github.com/inmaggp/Community-Detection-Problem-Based-on-Polarization-Measures.-An-application-to-Twitter-the-COVID-19-}}})). Nevertheless, for the clustering process, we focus on its weak component, which contains $261$ nodes and $484$ weighted edges. The obtained network, $G=(V,E)$, with adjacency matrix, $A$, is showed in Figure \ref{fig:network} (taking into consideration the weight of each edge, the degree of the nodes is represented by their size in the image, so that the bigger nodes will represent the users with the most amount of interactions).  Then, considering the membership degrees of each node to the poles $X_A$ (being a ``detractor'' of the Spanish government)  and $X_b$ (being a ``supporter'' of the Spanish government), represented by $\eta_{X_A}$ and $\eta_{X_B}$, respectively, we can calculate the Polarization measure $JDJ_{pol}$ (see \mbox{Definition \ref{def:JDJ}}) from which we define the matrix $P$, according to the Equation \eqref{eq:Pmas}. The membership degrees considered can be found in  GitHub$^1$. It provides us the non-polarization fuzzy measure $\mu_P$ which is one of the components of $\widetilde{G}=\left(V,E,\mu_P\right)$. {Note that the information provided by the non-polarization extended fuzzy graph goes further than that given by a crisp graph. It also includes the knowledge about the position of the nodes of $G$ with respect to an attitudinal axis, an information which cannot be modeled by classical tools.}

The measure $\mu_P$ depends on the selection of a negation operator, $N$, and two different types of aggregation operators: a grouping function $\varphi$ and an overlapping operator $\phi$.  As negation operator, we use $N(x)=\left(1-x\right)$. Concerning the aggregation operators, we use some of the most important operator in this field, having two different scenarios for the aggregation of the membership degrees: (a) $\phi=\min$ $ $ and $ $ $\varphi=\max$, and (b) $\phi=product$ $ $ and $ $ $\varphi=\max$.

Because of the characterization of $P$, and with $\mu_P$ being a fuzzy measure characterized as in Equation \eqref{eq:mupMas}, $P$ can be seen also as the adjacency matrix of $G_{\mu_P}$, $F$, so we can indistinctly consider both tools.

We apply the Polarization Louvain algorithm to find communities in the non-polarization extended fuzzy graph $\widetilde{G}=(V,E,\mu_P)$. Note that the obtained communities will be cohesive with the whole knowledge modeled by it, the structure of the graph as well as the additional information modeled by $\mu_P$. The notion of \textit{what is a community} will be closely connected with the aggregation operator $\theta$ chosen, as well as with the grouping operator $\varphi$ and the overlapping operator $\phi$. {Being able to consider the additional information when finding groups allow us to obtain realistic communities much more cohesive with the situation addressed, than those given by other methods which can not analyzed more information besides the structure of the graph.}
\begin{figure}[H]
    \includegraphics[scale=0.47]{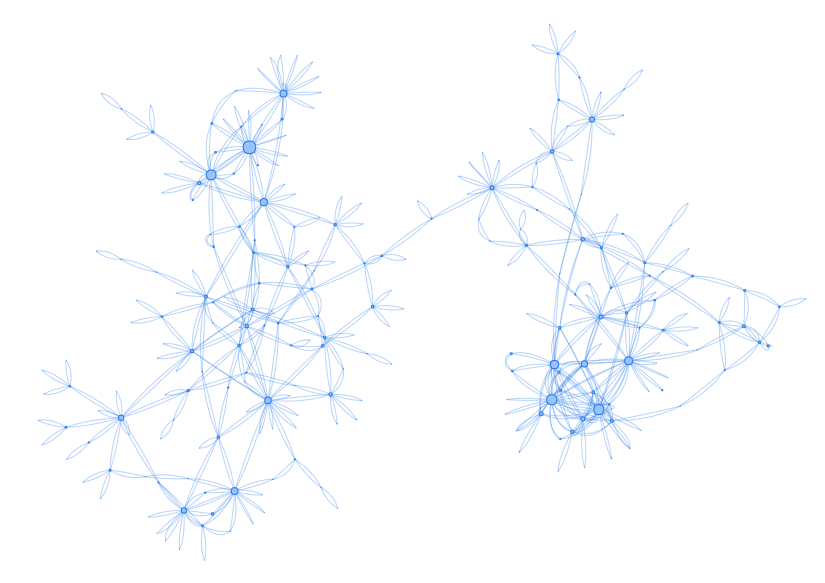}
    \caption{Graph $G=(V,E)$.}
    \label{fig:network}
\end{figure}
\vspace{-6pt}

\begin{figure}[H]
 \includegraphics[scale=0.4]{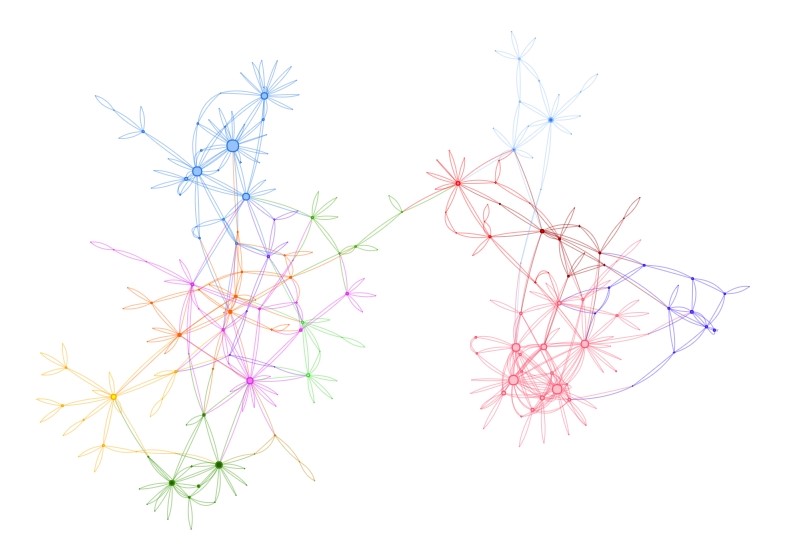}
    \caption{{Partition obtained} with the Louvain algorithm in the graph $G=(V,E)$.}  
    \label{fig:part_louvain}
\end{figure}

To combine the two components of $\widetilde{G}$,  we work with linear combinations of the matrices $A$ and $P$ assigning them an importance by means of the balancing parameter $\gamma\in[0,1]$, i.e., we consider the matrix $M=\theta(A,P)=\gamma A + (1-\gamma)P$.

The influence of each component of $\widetilde{G}$ varies depending on the value of $\gamma$. For values of $\gamma$ close to $1$, the structural component gains importance, so the groups contain nodes tightly connected in $A$. On the opposite, when $\gamma$ is close to $0$, the additional information modeled by $\mu_P$ turns decisive in the definition of the communities, so, if it is possible regarding the structure of $A$, the groups contain nodes with low Polarization level, i.e., nodes whose membership degree to each  pole is similar. In this case, those users about whom we can assume similar political viewpoint, will be in the same group.

We apply the Polarization Louvain algorithm for the two scenarios of grouping/ \linebreak overlapping functions previously mentioned, and considering the matrix $M=\gamma A + (1-\gamma)P$, for several values of the importance parameter, $\gamma=0.5, 0.4, 0.3, 0.2, 0.1, 0$. We also compute the Louvain algorithm with matrix $A$, on whose result, {showed in Figure} \ref{fig:part_louvain}, is based our comparison analysis. Note that the performance of the Louvain algorithm matches with the Polarization Louvain algorithm when $M=A$, ($\gamma=1$).

Here, we show how the organization of the groups keep changing depending on the importance of each component of $\widetilde{G}$ in the clustering process. Particularly, for the extreme cases, Louvain (in which there is no additional information) and $\gamma=0$ (in which the additional information gains all the importance), considering the two scenarios previously mentioned about the aggregation operators used. The results are shown in the Figures \ref{fig:part_alpha0_min} and \ref{fig:part_alpha0_prod}. In GitHub, we include a file in which we show the obtained partitions for every value of $\gamma$ considered, as well as the corresponding images.

\begin{figure}[H]

    \includegraphics[scale=0.4]{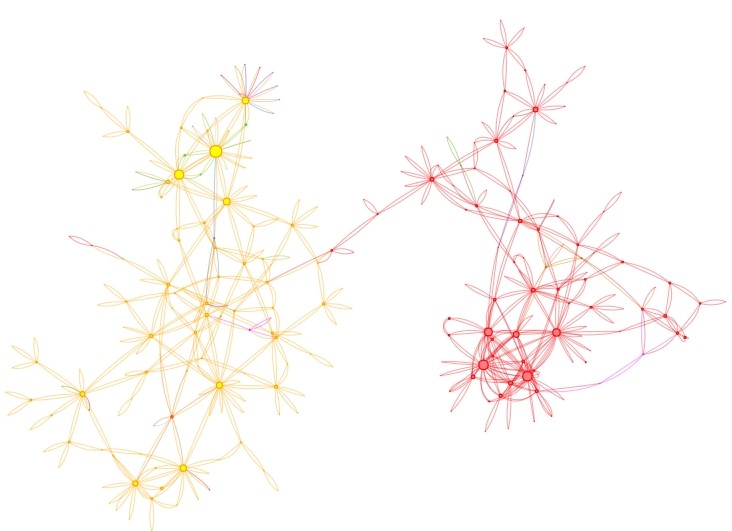}
    \caption{Partitions obtained with the Polarization Louvain algorithm in the non-polarization extended fuzzy graph  $\widetilde{G}=(V,E,\mu_P)$. $\gamma=0$; $\varphi=\max$; $\phi=\min$.}
    \label{fig:part_alpha0_min}
\end{figure}

\begin{figure}[H]

    \includegraphics[scale=0.4]{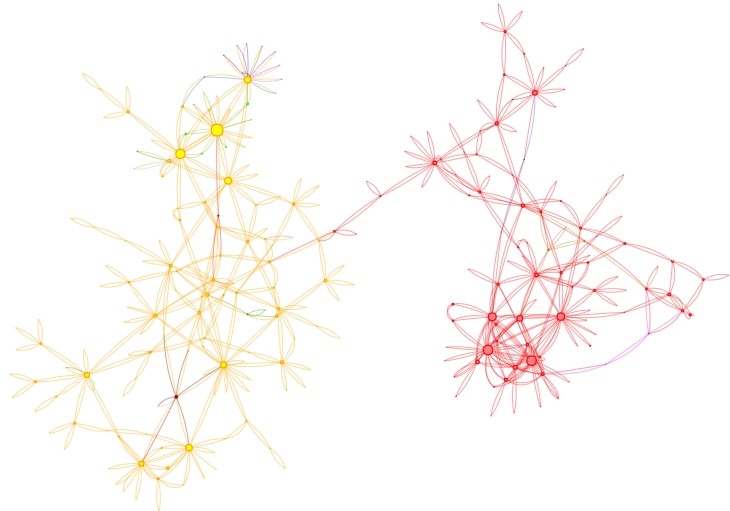}
    \caption{Partitions obtained with the Polarization Louvain algorithm in the non-polarization extended fuzzy graph $\widetilde{G}=(V,E,\mu_P)$. $\gamma=0$; $\varphi=\max$; $\phi=prod$.}
    \label{fig:part_alpha0_prod}
\end{figure}

Note how, when only the political viewpoint of the users is considered, the graph is divided into two main communities, so that we can easily differentiate between the detractors and the supporters of the Spanish government.

To measure the goodness of the obtained partitions, we refer to the $JDJ_{pol}$ measure. We agree on that a cohesive group should be composed by connected users with similar viewpoints. In this sense, we can say that a group is as cohesive as low is its corresponding $JDJ_{pol}$ value.

Note that the partitions obtained when consider several values of $\gamma$ vary in the number of communities. Then, to compare them, we consider the weighted average of the $JDJ_{pol}(C_i)$ value of all its communities. Thus, we calculate the Polarization value of the partition $P=\{C_1, \dots , C_s\}$ as

\begin{equation}\label{eq:jdjParti}
    pol(P) =  \Bigg[\frac{\sum_{i=1}^sJDJ_{pol}(C_i)*|C_i|}{\sum_{i=1}^s|C_i|} \Bigg]_{|C_i|>1}
\end{equation}

It is important to put attention on the fact that only the non-isolated communities will be considered to calculate $pol(P)$, i.e., groups with more  than one element  $|C_i|>1$; it does not make any sense consider how polarized is one element with respect itself.

In Tables \ref{ta:polMin} and \ref{ta:polProd}, we show the $JDJ_{pol}$ value of each community in the obtained partitions (only non-isolated communities), as well as the corresponding $pol(P)$. For each partition, we show the vector $(JDJ_{pol}(C_1), \dots , JDJ_{pol}(C_s))$, so that the $i$th component corresponds with $JDJ_{pol}(C_i)$. 
\end{paracol}
\nointerlineskip

\begin{specialtable}[H]
\widetable
\caption{Comparison of the obtained partitions. $\varphi=\max$ and $\phi=\min$.}
\label{ta:polMin}
\resizebox{\textwidth}{!}{%
\begin{tabular}{l c l c}
\toprule
\multicolumn{1}{l}{\begin{tabular}[l]{@{}l@{}}\boldmath{$\varphi=\max$}\\ \boldmath{$\phi=\min$}\end{tabular}} &
  \begin{tabular}[c]{@{}c@{}}$\#$ \textbf{Communities}\\ \boldmath{$|C_i|>1$}\end{tabular} &
  \multicolumn{1}{c}{\boldmath{$(JDJ(C_1), \dots , JDJ(C_s))$}} &
  \boldmath{$pol(P)$} \\ \midrule
$Louvain$ &
  14 &
  (0.256, 0.514, 0.253, 0.301, 0.458, 0.377, 0.302, 0.4403, 0.459, 0.349, 0.190, 0.475, 0.108, 0.415) & 0.359 \\ 
$\gamma = 0.5$ &
  11 &
  (0.239, 0.259, 0.513, 0.297, 0.377, 0.440, 0.514, 0.257, 0.459, 0.415, 0.455) & 0.341 \\ 
$\gamma = 0.4$ &
  8 &
  (0.254, 0.332, 0.259, 0.513, 0.450, 0.514, 0.459, 0.455) & 0.348
   \\ 
$\gamma = 0.3$ &
  7 &
  (0.304, 0.300, 0.253, 0.513, 0.512, 0.526, 0.246) & 0.343 \\ 
$\gamma = 0.2$ &
  8 &
  (0.334, 0.267, 0.444, 0.512, 0.462, 0.440, 0.528, 0.246) & 0.330 \\ 
$\gamma = 0.1$ &
  7 &
  (0.323, 0.273, 0.418, 0.482, 0.440, 0.462, 0.246) &
  0.319 \\ 
$\gamma = 0$ &
  8 &
  (0.302, 0.263, 0.439, 0.463, 0.277, 0.440, 0.462, 0.246) & 0.292 \\ \bottomrule
\end{tabular}%
}
\end{specialtable}
\vspace{-6pt}
\begin{specialtable}[H]
\widetable
\caption{Comparison of the obtained partitions. $\varphi=\max$ and $\phi=prod$.}
\label{ta:polProd}
\resizebox{\textwidth}{!}{%
\begin{tabular}{l c l c}
\toprule
\multicolumn{1}{l}{\begin{tabular}[l]{@{}l@{}}\boldmath{$\varphi=\max$}\\ \boldmath{$\phi=prod$}\end{tabular}} &
  \begin{tabular}[c]{@{}c@{}}\boldmath{$\#$} \textbf{Communities}\\ \boldmath{$|C_i|>1$}\end{tabular} &
  \multicolumn{1}{c}{\boldmath{$(JDJ(C_1), \dots , JDJ(C_s))$}} &
  \boldmath{$pol(P)$} \\ \midrule
$Louvain$   & 14 & (0.218, 0.454, 0.228, 0.261, 0.378, 0.296, 0.261p, 0.359, 0.389, 0.306, 0.168, 0.392, 0.102, 0.258) & 0.306 \\
$\gamma = 0.5$ & 11 & (0.220, 0.453, 0.190, 0.260, 0.261, 0.296, 0.359, 0.326, 0.382, 0.389, 0.258)   & 0.299 \\ 
$\gamma = 0.4$ & 9  & (0.214, 0.281, 0.220, 0.453, 0.363, 0.389, 0.369, 0.258, 0.343) & 0.292 \\
$\gamma = 0.3$ & 7  & (0.257, 0.251, 0.220, 0.453, 0.369, 0.417, 0.343)   & 0.292 \\ 
$\gamma = 0.2$ & 7  & (0.259, 0.228, 0.453, 0.369, 0.417, 0.249, 0.186)   & 0.289 \\ 
$\gamma = 0.1$ & 7  & (0.274, 0.228, 0.393, 0.417, 0.369, 0.249, 0.186) & 0.277 \\ 
$\gamma = 0$   & 8  & (0.256, 0.224, 0.316, 0.376, 0.199, 0.243, 0.249, 0.186) & 0.244 \\ \bottomrule
\end{tabular}%
}
\end{specialtable}

\begin{paracol}{2}
\switchcolumn

As it can be seen in previous tables, as well as in Figure \ref{fig:pol_alpha_op}, the $pol(P)$ value related to those partitions obtained with the Polarization Louvain algorithm is lower than the one related to the partition provided by the Louvain algorithm. Then, we can assert that this method provides more cohesive community structures according to the reality modeled. 
\begin{figure}[H]
    \includegraphics[scale=0.35]{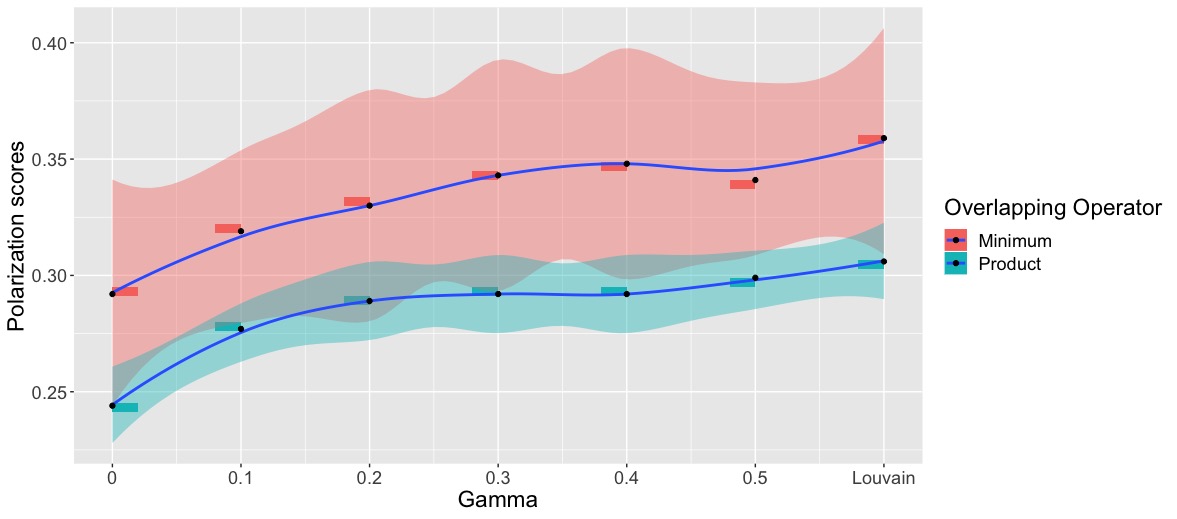}
    \caption{Polarization values of the partition $P=\{C_1, \dots , C_s\}$ by overlapping operators.}
    \label{fig:pol_alpha_op}
    \end{figure}

{To illustrate this, in the following figures we show an example of how two pairs of nodes which should belong to the same communities, respectively, are split into four different communities with the Louvain algorithm. On one hand, we have nodes ``38'' and ``115'', both left-wing political parties that teamed back in march 2019. On the other hand, we have nodes ``76'' , a right-wing political party, and ``203'', a member of this political group. After applying the Polarization Louvain algorithm, those pairs are clustered into the same communities (see Figure  \ref{fig13}a,b). Let us note that mentioned images are a zoom over the whole network, so not all the edges incident in these nodes are shown. Although it may seem that some nodes grouped in the same communities are not connected by edges (for example, nodes ``76'' and ``203'' in the image Figure  \ref{fig13}b) all of them are properly connected in the network.}

\begin{figure}[H]
\begin{subfigure}{0.45\linewidth}
\centering
    \includegraphics[scale=0.16]{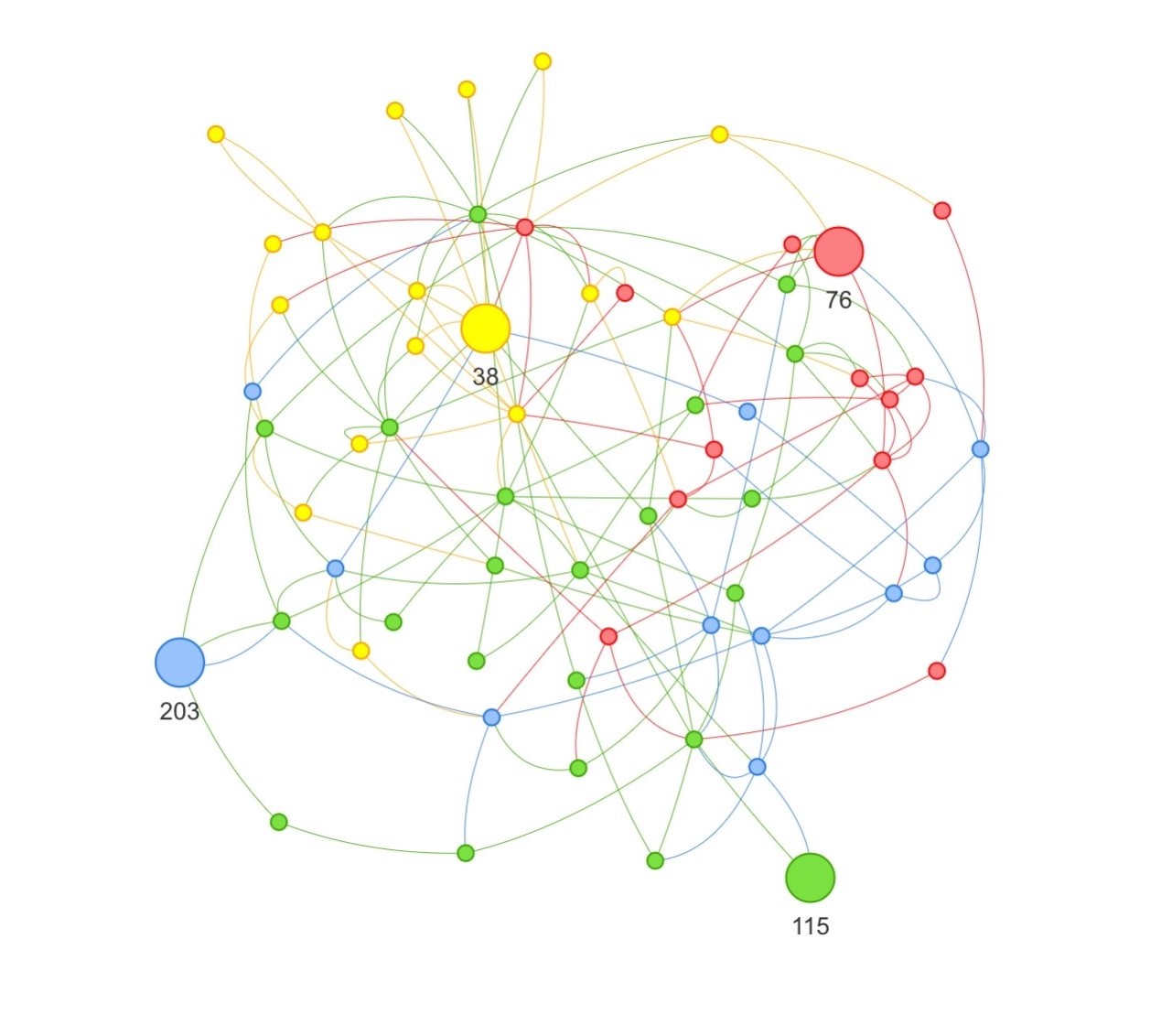}
    \caption{{Nodes ``38'', ``76'', ``115'', and ``203'' grouped by  Louvain algorithm.}}
    \label{fig:lou_comp}
\end{subfigure}
    \begin{subfigure}{0.53\linewidth}

    \includegraphics[scale=0.16]{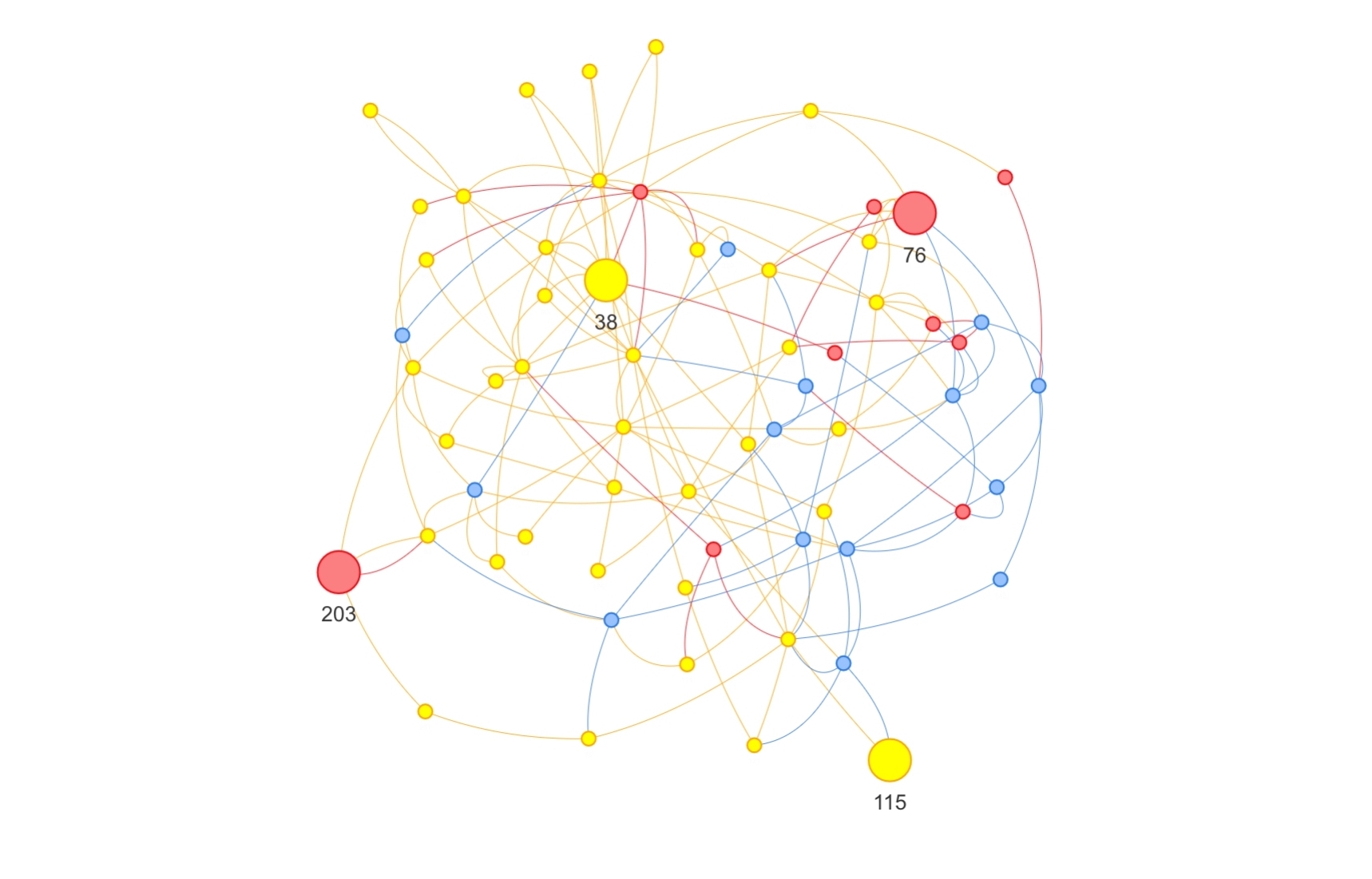}\vspace{0.55mm}
    \caption{{Nodes ``38'', ``76'', ``115'', and ``203'' grouped by \\ Polarization Louvain algorithm.}}
    \label{fig:alpha_comp}
\end{subfigure}
\caption{Zoom of the whole network.}
    \label{fig13}
    \end{figure}

\section{Discussion}\label{sec:discussion}
There are several points to be discussed at this end.

From a theoretical point of view, it is undeniable that complex models fix  the reality better than classical tools. Having several criteria to be considered makes the resolution process of a problem more complex, but it is certainly worth it.

Classically, the methods proposed to find communities in a graph only analyze its structural features. Far from this assumption, in this work we have taken into consideration several aspects inherent to reality, which, with a proper process of modeling and analysis, can be considered as different criteria in the community detection problem.

Then, we distinguish between different types of information. On the one hand, we deem the crisp knowledge which could be easily related to the classical graphs considered in the literature. This type of knowledge is unalterable and objective, in the sense that it exists and no changes could be made about it. It is the case of the direct connections represented by the edges of a graph. Particularly, in the real case here addressed  based on the Online Social Network Twitter, we have worked with the retweet (RT) network. It is composed of objective information directly obtained from the social site. On the other hand, we analyze other types of information sources inherent to the people who discuss in Twitter. A wide range of different aspects could be considered, from the factual issues related to objective knowledge about the people as, for example, the distance that separates them or the common followers, to the more subjective points related to ideology or feelings.

In the context of great political instability enhanced by the global COVID-19 crisis, we agree on the importance of analyzing the political position of several people who are highly influential on Twitter. The study of feelings, ideology, and political principles, is always a hard matter in which many inaccurate details have to be taken into account. To deal with the vagueness and vagueness related to the analysis of political attitudes, we work with fuzzy measures, added in the modeling process to the crisp graph. In this vein, we work with the  non-polarization extended fuzzy graph $\widetilde{G}=(V,E,\mu_P)$, where $G=(V,E)$ is the weighted graph which represents the RT network, and $\mu_P$ is a fuzzy measure which defines relations between the elements of $V$, depending on their position in a political axis. To define this fuzzy measure, we consider the $JDJ_{pol}$ measure, which quantifies the Polarization of a given society.
  
Several criteria have to be fixed for the calculation of $JDJ_{pol}$, as, for example, the aggregation operator $\phi$ and the grouping function $\varphi$. In this paper, we have selected considered some of the most popular functions in this field, specifically, $\varphi=\max$ $ $ and  $ $ $\phi\in \{ \min, product \}$. Note that these operators play an essential role in the value of the $JDJ_{pol}$ measure (and also the negation operator $N$ if we are interested in considering the opposite of $JDJ_{pol}$) and thus in the community detection problem here addressed. Therefore, it would be interesting to analyze how the structure of the partitions keeps changing according to the operators considered. 

In the same manner, the operator $\theta$ involved in the Polarization Louvain algorithm impacts on the community structure detected, in terms of how to aggregate both components of a non-polarization extended fuzzy graph  $\widetilde{G}=(V,E,\mu_P)$ (the structure and the closeness between the nodes). We agree on considering linear combinations of the two matrices involved, in order to assign an ``importance'' to each of them, by means of a balancing or weighting factor $\gamma\in [0,1]$. This procedure allows us to examine the changes that occurs in the structure of communities, according to the how much influence on its definition each of these components. Then, considering the aggregation $\gamma A + (1-\gamma)P$, those values of $\gamma$ which are close to $1$ are related to partitions in which the nodes of the same group are densely connected in $G$, whereas for lower values of $\gamma$ it is important to maintain together nodes with high values of closeness (without omitting the structure of $G$).

Regarding Polarization values, the Louvain algorithm shows the highest values of Polarization, as it can be seen in the Tables \ref{ta:polMin} and \ref{ta:polProd}, as well as in the Figure \ref{fig:pol_alpha_op}. In this work, we propose a new method for community detection, which in our opinion has strong theoretical and applied connotations. The extra-information provided by  the measure of Polarization $JDJ_{pol}$ matches up with community detection algorithms due to their close conceptual relationship. The fact of adding  Polarization scores implies taking into account the similarity between individuals along an attitudinal axis. In this vein, having new information closely related to the purposes for which the community detection algorithms are applied, makes the communities more cohesive with a greater homogeneity degree, so that this construction of the communities fixes better the reality. In our case, the aim is to cluster the nodes according to their position towards the Spanish government.

\section{Conclusions}\label{sec:conclusions}
In this paper, we work in the definition of a polarization fuzzy measure obtained from a Polarization measure. It is a model to represent the capacity of a set of elements to argue. Then, we introduce a new tool which combines the capacity represented by that polarization fuzzy measure, with the connections between elements modeled by a graph:  the polarization extended fuzzy graph.

In order to handle situations in which the interest is not in the capacity of the elements to argue, but it is in their capacity to peacefully dialogue, we suggest the definition of the non-polarization fuzzy measure. Similarly as it is proposed concerning the polarization fuzzy measure, we introduce the non-polarization extended fuzzy graph, which allows the representation of the capacity to dialogue of a set of elements combined with their connections throughout a graph.

Then, we address the community detection problem in an extended context regarding the existence of several criteria to be taken into account. On the one hand, we consider the representation of the direct connections between the individuals represented by a crisp network $G=(V,E)$. On the other hand, we know the position of all the elements (represented by the nodes) in any attitudinal axis, information not inherent to the structural representation of their connections.

From this extra-information, understood as the membership degree of each element to two extreme poles, the $JDJ_{pol}$ polarization measure is defined, which will be the base of characterization of a non-polarization fuzzy measure $\mu_P$. Then, we define the non-polarization extended fuzzy graph $\widetilde{G}=(V,E,\mu_P)$, on which we set the basis of the community detection problem based on fuzzy measures.
On this assumption, we address a real case obtained from Twitter.

The graphic representation of a network reflects the structure of a given set of nodes according to their interactions and behavior. From this point of view, the sociological phenomenon which drive all these interactions is called homophily \cite{homophily}. According to the concept of homophily, a set of individuals or nodes are grouped and interact with each other according to their similarities. So that the concept of Polarization is emanated from homophily and, more specifically, homogeneity \cite{robles2019polarizacion}, appearing in those scenarios in which a set of nodes or individuals are split into two opposite groups. In this vein, the measurement of Polarization provides the adequate clues for community detection problems. Furthermore, the fuzzy-set theoretical approach provides the appropriate resources in order to tackle this issue from a realistic position. Adding the extra-information provided by $JDJ_{pol}$ has a double benefit: (1) it not only allows increasing the homogeneity degree intra-community, but (2) it also provides essential information in those cases where there are some nodes with a non-clear membership with the classical community detection algorithms. Furthermore, note the importance of the aims and hypothesis of the study which should be the same for both, community detection application, and Polarization measurement. Thus, the synergy between community detection algorithm and other measures will be an optimal solution. As a consequence, not only the integration of a given community is more realistic but also the global topographic structure.

{Regarding the construction of the membership degree functions, they can be constructed by different approaches as well as they reflect the proximity of a given individual to the poles. In \cite{guevara2020measuring}, the authors proposed a triangular membership function where each of the categories used in their example---they apply the measure to a categorical variable---had a given probability of belonging to each pole assuming that the lowest value is one pole and highest is the other. In this case, we use as membership functions the support vector machine classifier outputs, which give us soft information that can be used to know how close each item is to the classes $X_A$ and $X_B$. From this soft value we build  the membership function of each individual to each pole, obtaining $\eta_{X_A}(i)$ and $\eta_{X_B}(i)$ by which we can know the degree in which a given individual belong to both poles.}

    {To conclude this final section, we would like to mention some points. One of the most difficult problems in fuzzy sets theory is how to build a membership function. In this work, we need to build $\eta_{X_A}$ and $\eta_{X_B}$, in order to then build the polarization fuzzy measure. In \cite{guevara2020measuring}, thee authors face the construction of the membership from a fuzzy sets perspective. Nevertheless,  it is not the main objective of this paper. In this work, we apply machine learning algorithms which allow us to measure the closeness of each node to each pole. From this information, we can build the membership function of each individual to each pole. Note that this procedure could be replicated easily to other similar situations in which we had the knowledge of some items to the two poles and we apply machine learning techniques to build the membership functions.}


\authorcontributions{Conceptualization, I.G., J.A.G., and D.G.; methodology, I.G., J.A.G. and D.G; software, I.G. and J.A.G.; validation, D.G., J.C., and R.E.; formal analysis, I.G. and J.A.G.; investigation, I.G. and J.A.G.; resources, I.G., J.A.G., D.G., J.C., and R.E.; data curation, I.G. and J.A.G.; writing---original draft preparation, I.G. and J.A.; writing---review and editing, I.G., J.A.G., D.G., J.C., and R.E.; visualization, I.G. and J.A.G.; supervision, D.G., J.C., and R.E.; project administration, D.G., J.C., and R.E.; funding acquisition, D.G., J.C., and R.E. All authors have read and agree to the published version of the manuscript.}

\funding{This research has been partially supported by the Government of Spain, Grant Plan Nacional de I+D+i, MTM2015-70550-P, PGC2018096509-B-I00, TIN2015-66471-P and PID2019-106254RB-I00, and the CT17/17-CT18/17.}

\institutionalreview{Not applicable}

\informedconsent{Not applicable}

\dataavailability{MDPI Research Data Policies at {https://www.mdpi.com/ethics}.}

\conflictsofinterest{The authors declare no conflicts of interest. The funders had no role in the design of the study; in the collection, analyses, or interpretation of data; in the writing of the manuscript; or in the decision to publish the results.} 

\abbreviations{The following abbreviations are used in this manuscript:\\

\noindent 
\begin{tabular}{@{}ll}
SNA & Social Networks Analysis
\end{tabular}}

\appendixtitles{no} 

\end{paracol}
\reftitle{References}

\begin{thebibliography}{999}

\bibitem[Bennett \em{et~al.}(2015)Bennett, Kittas, Muirhead, Papageorgiou, and
  Tsoka]{bennett}
Bennett, L.; Kittas, A.; Muirhead, G.; Papageorgiou, L.; Tsoka, S.
\newblock Detection of composite communities in multiplex biological networks.
\newblock {\em Sci. Rep.} {\bf 2015}, {\em 5},~10345.

\bibitem[Chaker \em{et~al.}(2017)Chaker, Al~Aghbari, and Junejo]{chaker}
Chaker, R.; Al~Aghbari, Z.; Junejo, I.
\newblock Social network model for crowd anomaly detection and localization.
\newblock {\em Pattern Recognit.} {\bf 2017}, {\em 61},~266--281.

\bibitem[Harakawa \em{et~al.}(2014)Harakawa, Ogawa, and Haseyama]{harakawa}
Harakawa, R.; Ogawa, T.; Haseyama, M.
\newblock Accurate and efficiet extration of hierarchical structure of web
  communities for web video retrieval.
\newblock {\em ITE Trans. Media Technol. Appl.s} {\bf
  2014}, {\em 2},~287--297.

\bibitem[Tamura \em{et~al.}(2015)Tamura, Kobayashi, and Ihara]{tamura}
Tamura, K.; Kobayashi, Y.; Ihara, Y.
\newblock Evolution of individual versus social learning on social networks.
\newblock {\em J. R. Soc. Interface} {\bf 2015}, {\em
  12},~20141285.

\bibitem[Esteban and Ray(1994)]{ER94}
Esteban, J.M.; Ray, D.
\newblock On the measurement of polarization.
\newblock {\em Econom. J. Econom. Soc.} {\bf 1994},
  {\em 62},~819--851.

\bibitem[Guevara \em{et~al.}(2020)Guevara, G{\'o}mez, Robles, and
  Montero]{guevara2020measuring}
Guevara, J.A.; G{\'o}mez, D.; Robles, J.M.; Montero, J.
\newblock Measuring Polarization: A Fuzzy Set Theoretical Approach.
\newblock  In Proceedings of the International Conference on Information Processing and Management of
  Uncertainty in Knowledge-Based Systems, Lisbon, Portugal, 15--19 June   2020; Springer: Berlin, Germany,  2020; pp. 510--522.

\bibitem[Clauset \em{et~al.}(2004)Clauset, Newman, and Moore]{clauset}
Clauset, A.; Newman, M.; Moore, C.
\newblock Finding community structure in very large networks.
\newblock {\em Phys. Rev. E} {\bf 2004}, {\em 70},~066111.

\bibitem[Newman(2012)]{large}
Newman, M.
\newblock Communities, modules and large-scale structure in networks.
\newblock {\em Phys. Rev.} {\bf 2012}, {\em 8},~25--31.

\bibitem[Girvan and Newman(2002)]{biological}
Girvan, M.; Newman, M.
\newblock Community structure in social and biological networks.
\newblock {\em Proc. Natl. Acad. Sci.  USA} {\bf 2002},
  {\em 99},~7821--7826.

\bibitem[Blondel \em{et~al.}(2008)Blondel, Guillaume, Lambiotte, and
  Lefevre]{blondel}
Blondel, V.; Guillaume, J.; Lambiotte, R.; Lefevre, E.
\newblock Fast unfolding of communities in large networks.
\newblock {\em J. Stat.-Mech. Theory Exp.} {\bf
  2008}, {\em 10}.

\bibitem[Waltman and Van~Eck(2013)]{smart}
Waltman, L.; Van~Eck, N.
\newblock A smart local moving algorithm for large-scale modularity-based
  community detection.
\newblock {\em Eur. Phys. J. B} {\bf 2013}, {\em 86},~473.

\bibitem[Newman and Girvan(2004)]{girvanNewman}
Newman, M.; Girvan, M.
\newblock Finding and evaluating community structure in networks.
\newblock {\em Phys. Rev. E} {\bf 2004}, {\em 69}.

\bibitem[Fortunato(2010)]{fortunato}
Fortunato, S.
\newblock Community detection in graphs.
\newblock {\em Phys. Rep.-Rev. Sect. Phys. Lett.} {\bf 2010},
  {\em 486},~75--174.

\bibitem[G\'omez \em{et~al.}(2008)G\'omez, Gonz\'alez-Arang\"uena, Manuel,
  Owen, and Saboy\'a]{cohesiveness}
G\'omez, D.; Gonz\'alez-Arang\"uena, E.; Manuel, C.; Owen, G.; Saboy\'a, M.
\newblock The cohesiveness of subgroups in social networks: A view from game
  theory.
\newblock {\em Ann. Oper. Res.} {\bf 2008}, {\em 158},~33--46.

\bibitem[G\'omez \em{et~al.}(2003)G\'omez, Gonz\'alez-Arang\"uena, Manuel,
  Owen, del Pozo, and Tejada]{centralityPower}
G\'omez, D.; Gonz\'alez-Arang\"uena, E.; Manuel, C.; Owen, G.; del Pozo, M.;
  Tejada, J.
\newblock Centrality and power in social networks: A game theoretic approach.
\newblock {\em Math. Soc. Sci.} {\bf 2003}, {\em 46},~27--54.

\bibitem[Devarajan \em{et~al.}(2019)Devarajan, Fatima, Vairavasundaram, and
  Ravi]{Devarajan}
Devarajan, M.; Fatima, N.; Vairavasundaram, S.; Ravi, L.
\newblock Swarm intelligence clustering ensemble based point of interest
  recommendation for social cyber-physical systems.
\newblock {\em J. Intell. Fuzzy Syst.} {\bf 2019}, {\em
  36},~4349--4360.

\bibitem[Nair and Sarasamma(2007)]{Nair2007}
Nair, P.; Sarasamma, S.
\newblock Data mining through fuzzy social network analysis.
\newblock { In Proceedings of the NAFIPS 2007---2007 Annual Meeting of the North American Fuzzy
  Information Processing Society},  San Diego, CA, USA, 24--27 June { 2007}; pp. 251--255.

\bibitem[Guti\'errez \em{et~al.}(2020{\natexlab{a}})Guti\'errez, G\'omez,
  Castro, and Esp\'inola]{infus}
Guti\'errez, I.; G\'omez, D.; Castro, J.; Esp\'inola, R.
\newblock A New Community Detection Algorithm Based on Fuzzy Measures.
\newblock  In \emph{Advances in Intelligent Systems and Computing Series, Proceedings of the Intelligent
  and Fuzzy Techniques in Big Data Analytics and Decision Making  INFUS 2019, San Diego, CA, USA, 24--27 June 2020}; Kahraman, C.,
  Cebi, S., Cevik~Onar, S., Oztaysi, B., Tolga, A., Sari, I., Eds.; 
  Springer: Cham, Switzerland, 2020; Volume 1029, pp. 133--140.

\bibitem[Guti\'errez \em{et~al.}(2020{\natexlab{b}})Guti\'errez, G\'omez,
  Castro, and Esp\'inola]{ampliinfus}
Guti\'errez, I.; G\'omez, D.; Castro, J.; Esp\'inola, R.
\newblock Fuzzy Measures: A solution to deal with community detection problems
  for networks with additional information.
\newblock {\em J. Intell. Fuzzy Syst.} {\bf 2020}, {\em
  39},~6217--6230.
{\detokenize{10.3233/JIFS-18909}}.

\bibitem[Guti\'errez \em{et~al.}(2020{\natexlab{c}})Guti\'errez, G\'omez,
  Castro, and Esp\'inola]{multipleBip}
Guti\'errez, I.; G\'omez, D.; Castro, J.; Esp\'inola, R.
\newblock Multiple bipolar fuzzy measures: An application to community
  detection problems for networks with additional information.
\newblock {\em Int. J. Comput. Intell. Syst.}
  {\bf 2020}, {\em 13},~1636--1649.

\bibitem[Rosenfeld(1975)]{rosenfeldFG}
Rosenfeld, A.
\newblock Fuzzy Graphs.
\newblock {\em Fuzzy Sets Their Appl.} {\bf 1975}, pp. 77--95.

\bibitem[Zadeh(1965)]{zadeh1965}
Zadeh, L.
\newblock Fuzzy sets.
\newblock {\em Information and Control} {\bf 1965}, {\em 8},~338--353.

\bibitem[Yaqoob \em{et~al.}(2019)Yaqoob, Gulistan, Kadry, and Wahab]{yaqoob}
Yaqoob, N.; Gulistan, M.; Kadry, S.; Wahab, H.
\newblock Complex Intuitionistic Fuzzy Graphs with Application in Cellular
  Network Provider Companies.
\newblock {\em Mathematics} {\bf 2019}, {\em 7},~35.

\bibitem[Zuo \em{et~al.}(2019)Zuo, Pal, and Dey]{picturefuzzy}
Zuo, C.; Pal, A.; Dey, A.
\newblock New Concepts of Picture Fuzzy Graphs with Application.
\newblock {\em Mathematics} {\bf 2019}, {\em 7},~470.

\bibitem[Mordeson and Nair(2000)]{mordeson}
Mordeson, J.; Nair, P.
\newblock Fuzzy Graphs and Fuzzy Hypergraphs.
\newblock {\em Stud. Fuzziness Soft Comput.} {\bf 2000}, {\em
  46},~19--81.

\bibitem[Beliakov(2020)]{beliakov20}
Beliakov, G.
\newblock On random generation of supermodular capacities.
\newblock {\em IEEE Trans. Fuzzy Syst.} {\bf 2020}.

\bibitem[Sugeno(1977)]{sugeno}
Sugeno, M.
\newblock Fuzzy measures and fuzzy integrals: A survey.
\newblock {\em Fuzzy Autom. Decis. Process.} {\bf 1977}, {\em
  78},~89--102.

\bibitem[Newman(2006)]{newman2006}
Newman, M.
\newblock Modularity and community structure in networks.
\newblock {\em Proc. Natl. Acad. Sci. USA} {\bf 2006},
  {\em 103},~8577--8582.

\bibitem[Reynal-Querol(2001)]{reynal2001ethnic}
Reynal-Querol, M.
\newblock Ethnic and Religious Conflicts, Political Systems and Growth.
\newblock Ph.D. Thesis, London School of Economics and Political Science.
  University of London, London, UK, 2001.

\bibitem[Apouey(2007)]{Apouey_2007}
Apouey, B.
\newblock Measuring health polarization with self‐assessed health data.
\newblock {\em Health Econ.} {\bf 2007}, {\em 16},~20.

\bibitem[Permanyer and D'Ambrosio(2015)]{permanyer2015measuring}
Permanyer, I.; D'Ambrosio, C.
\newblock Measuring social polarization with ordinal and categorical data.
\newblock {\em J. Public Econ. Theory} {\bf 2015}, {\em
  17},~311--327.

\bibitem[G{\'o}mez \em{et~al.}(2016)G{\'o}mez, Rodriguez, Montero, Bustince,
  and Barrenechea]{gomez2016n}
G{\'o}mez, D.; Rodriguez, J.T.; Montero, J.; Bustince, H.; Barrenechea, E.
\newblock n-Dimensional overlap functions.
\newblock {\em Fuzzy Sets Syst.} {\bf 2016}, {\em 287},~57--75.

\bibitem[Bustince \em{et~al.}(2011)Bustince, Pagola, Mesiar, Hullermeier, and
  Herrera]{bustince2011grouping}
Bustince, H.; Pagola, M.; Mesiar, R.; Hullermeier, E.; Herrera, F.
\newblock Grouping, overlap, and generalized bientropic functions for fuzzy
  modeling of pairwise comparisons.
\newblock {\em IEEE Trans. Fuzzy Syst.} {\bf 2011}, {\em
  20},~405--415.

\bibitem[Grabisch(1997)]{gra2}
Grabisch, M.
\newblock $k$-order additive discrete fuzzy measures and their representation.
\newblock {\em Fuzzy Sets Syst.} {\bf 1997}, {\em 92},~167--189.

\bibitem[Guimera and Amaral(2005)]{guimera}
Guimera, R.; Amaral, L.
\newblock Functional cartography of complex metabolic networks.
\newblock {\em Nature} {\bf 2005}, {\em 433},~895--900.

\bibitem[Flake \em{et~al.}(2002)Flake, Lawrence, Giles, and
  Coetzee]{selfOrganization}
Flake, G.; Lawrence, S.; Giles, C.; Coetzee, F.
\newblock Self-organization and identification of web communities.
\newblock {\em Computer} {\bf 2002}, {\em 35},~66--70.

\bibitem[Zou \em{et~al.}(2017)Zou, Chen, Li, Lu, and Lin]{discreteBack}
Zou, F.; Chen, D.; Li, S.; Lu, R.; Lin, M.
\newblock Community detection in complex networks: Multi-objective discrete
  backtracking search optimization algorithm with decomposition.
\newblock {\em Appl. Soft Comput.} {\bf 2017}, {\em 53},~285--295.

\bibitem[Liu \em{et~al.}(2020)Liu, Wang, and Liu]{penalized}
Liu, J.; Wang, J.; Liu, B.
\newblock Community Detection of Multi-Layer Attributed Networks via Penalized
  Alternating Factorization.
\newblock {\em Mathematics} {\bf 2020}, {\em 8},~239.

\bibitem[Gupta \em{et~al.}(2017)Gupta, Mittal, Gupta, Singhal, Gupta, and
  Kumar]{guptaParallel}
Gupta, S.; Mittal, S.; Gupta, T.; Singhal, I.; Gupta, A.; Kumar, N.
\newblock Parallel quantum-inspired evolutionary algorithms for community
  detection in social networks.
\newblock {\em Appl. Soft Comput.} {\bf 2017}, {\em 61},~331--353.

\bibitem[Vitali and Battiston(2013)]{vitalli}
Vitali, S.; Battiston, S.
\newblock The Community Structure of the Global Corporate Network.
\newblock {\em SSNR Electron. J.} {\bf 2013}, {\em 8}, doi:10.2139/ssrn.2198974.

\bibitem[Carnivali \em{et~al.}(2020)Carnivali, Vieira, Ziviani, and
  Esquef]{covec}
Carnivali, G.; Vieira, A.; Ziviani, A.; Esquef, P.
\newblock Co\MakeUppercase{v}e\MakeUppercase{c}: Coarse-Grained Vertex
  Clustering for Efficient Community Detection in Sparse Complex Networks.
\newblock {\em Inf. Sci.} {\bf 2020}, {\em 522},~180--192.

\bibitem[Riolo and Newman(2020)]{consistencyRiolo}
Riolo, M.; Newman, M.
\newblock Consistency of community structure in complex networks.
\newblock {\em Phys. Rev. E} {\bf 2020}, {\em 101},~052306.

\bibitem[de~Blas \em{et~al.}(2018)de~Blas, Martin, and Gomez]{de2018combined}
de~Blas, C.S.; Martin, J.S.; Gomez, D.
\newblock Combined social networks and data envelopment analysis for ranking.
\newblock {\em Eur. J. Oper. Res.} {\bf 2018}, {\em
  266},~990--999.

\bibitem[Guti\'errez \em{et~al.}()Guti\'errez, G\'omez, Castro, and
  Esp\'inola]{escim19}
Guti\'errez, I.; G\'omez, D.; Castro, J.; Esp\'inola, R.
\newblock A new community detection problem based on bipolar fuzzy measures.
\newblock {\em Stud. Comput. Intell.} \textbf{2021},
\newblock \uppercase{I}n \uppercase{P}ress.

\bibitem[Guti\'errez \em{et~al.}(2020)Guti\'errez, G\'omez, Castro, and
  Esp\'inola]{wcci2020}
Guti\'errez, I.; G\'omez, D.; Castro, J.; Esp\'inola, R.
\newblock Fuzzy Sugeno $\lambda$-Measures and Theirs Applications to Community
  Detection Problems.  In Proceedings of the IEEE International Conference on Fuzzy Systems, Glasgow, UK,  19--24   July 2020;
  pp. 1--6.

\bibitem[Barroso \em{et~al.}(2020)Barroso, Guti{\'e}rrez, G{\'o}mez, Castro,
  and Esp{\'i}nola]{ipmu2020}
Barroso, M.; Guti{\'e}rrez, I.; G{\'o}mez, D.; Castro, J.; Esp{\'i}nola, R.
\newblock Group Definition Based on Flow in Community Detection.
\newblock In \emph{Information Processing and Management of Uncertainty in
  Knowledge-Based Systems}; Lesot, M.J., Vieira, S., Reformat, M.Z., Carvalho,
  J.P., Wilbik, A., Bouchon-Meunier, B., Yager, R.R., Eds.; Springer
  International Publishing: Cham,  Switzerland, 2020; pp. 524--538.

\bibitem[Guti{\'e}rrez \em{et~al.}(2020)Guti{\'e}rrez, Barroso, G{\'o}mez,
  Castro, and Esp{\'i}nola]{flins2020}
Guti{\'e}rrez, I.; Barroso, M.; G{\'o}mez, D.; Castro, J.; Esp{\'i}nola, R.
\newblock Pattern-based clustering problem based on fuzzy measures.
\newblock {\em Dev. Artif. Intell. Technol. Comput. Robot.} {\bf 2020}, {\em 12},~412--420.

\bibitem[Shapley(1953)]{shapley}
Shapley, L.
\newblock A value for $n$-person games.
\newblock {\em Contribute. Theory Games} {\bf 1953}, {\em 2},~307--317.

\bibitem[Grabisch \em{et~al.}(1995)Grabisch, Nguyen, and Walker]{uncertainty}
Grabisch, M.; Nguyen, H.; Walker, E.
\newblock {\em Fundamentals of Uncertainty Calculi with Applications to Fuzzy
  Inference}; Kluwer Academic: Dordrecht,  The Netherlands, 1995.

\bibitem[Castro \em{et~al.}(2017)Castro, G\'omez, Molina, and Tejada]{imprSh}
Castro, J.; G\'omez, D.; Molina, E.; Tejada, J.
\newblock Improving polynomial estimation of the \uppercase{S}hapley value by
  stratified random sampling with optimum allocation.
\newblock {\em Comput. Oper. Res.} {\bf 2017}, {\em
  82},~108--188.

\bibitem[Castro \em{et~al.}(2009)Castro, G\'omez, and Tejada]{poliSh}
Castro, J.; G\'omez, D.; Tejada, J.
\newblock Polynomial calculation of the Shapley value based on sampling.
\newblock {\em Comput. Oper. Res.} {\bf 2009}, {\em
  36},~1726--1730.

\bibitem[Robles \em{et~al.}(2019)Robles, Atienza, G{\'o}mez, and
  Guevara]{robles2019polarizacion}
Robles, J.M.; Atienza, J.; G{\'o}mez, D.; Guevara, J.A.
\newblock La polarizaci{\'o}n de “La Manada”. El debate p{\'u}blico en
  Espa{\~n}a y los riesgos de la comunicaci{\'o}n pol{\'\i}tica digital.
\newblock {\em Tempo Soc.} {\bf 2019}, {\em 31},~193--216.

\bibitem[Kearney(2019)]{rtweet-package}
Kearney, M.W.
\newblock rtweet: Collecting and analyzing Twitter data.
\newblock {\em J. Open Source Softw.} {\bf 2019}, {\em 4},~1829.

\bibitem[Wang \em{et~al.}(2006)Wang, Sun, Zhang, and Li]{wang2006optimal}
Wang, Z.Q.; Sun, X.; Zhang, D.X.; Li, X.
\newblock An optimal SVM-based text classification algorithm.
\newblock   In Proceedings of the  2006 IEEE International Conference on Machine Learning and Cybernetics, Dalian, China, 13--16 August 2006; pp. 1378--1381.

\bibitem[Meyer \em{et~al.}(2020)Meyer, Dimitriadou, Hornik, Weingessel, and
  Leisch]{Re1071}
Meyer, D.; Dimitriadou, E.; Hornik, K.; Weingessel, A.; Leisch, F.
\newblock {\em e1071: Misc Functions of the Department of Statistics,
  Probability Theory Group (Formerly: E1071), TU Wien}; R package version 1.7-4;  {2020}. 

 

\bibitem[Huang and Ling(2005)]{AUC}
Huang, J.; Ling, C.
\newblock Using \uppercase{auc} and accuracy in evaluating learning algorithms.  \emph{IEEE Trans. Knowl. Data Eng. }
  {\bf 2005}.
\newblock {\em 17},~299--310.

\bibitem[Joachims(1998)]{joachim}
Joachims, T.
\newblock Text categorization with Support Vector Machines: Learning with many
  relevant features.
\newblock In  \emph{Machine Learning: ECML-98}; N{\'e}dellec, C., Rouveirol, C., Eds.;
  Springer: Berlin/Heidelberg,  Germany, 1998; pp. 137--142.

\bibitem[Park \em{et~al.}(2020)Park, Yoon, Lee, and Kim]{networkVisualization}
Park, J.; Yoon, S.; Lee, C.; Kim, J.
\newblock A Simple Method for Network Visualization.
\newblock {\em Mathematics} {\bf 2020}, {\em 8},~1020.

\bibitem[{Almende B.V.} \em{et~al.}(2019){Almende B.V.}, Thieurmel, and
  Robert]{visNetwork}
{Almende B.V.}.; Thieurmel, B.; Robert, T.
\newblock {\em visNetwork: Network Visualization using `vis.js' Library}; R package version 2.0.9;
 { 2019.}

\bibitem[McPherson \em{et~al.}(2001)McPherson, Smith-Lovin, and
  Cook]{homophily}
McPherson, M.; Smith-Lovin, L.; Cook, J.
\newblock Birds of a feather: Homophily in social networks.
\newblock {\em Annu. Rev. Sociol.} {\bf 2001}, {\em 27},~415--444.
\end{thebibliography}



\end{document}